\documentclass{article} % For LaTeX2e
\usepackage[preprint]{colm2026_conference}

\usepackage{microtype}
\usepackage{hyperref}
\usepackage{url}
\usepackage{booktabs}
\usepackage{graphicx}
\usepackage{booktabs}
\usepackage{multirow}
\usepackage[table]{xcolor}
\usepackage{amssymb}
\usepackage{amsmath}
\usepackage[inline]{enumitem}
\usepackage[skip=2pt]{caption}

\usepackage{lineno}

\definecolor{darkblue}{rgb}{0, 0, 0.5}

\hypersetup{colorlinks=true, citecolor=darkblue, linkcolor=darkblue, urlcolor=darkblue}
\newcommand{\header}[1]{\vspace*{1mm}\noindent\textbf{#1}.}

\newcommand{\cmark}{\checkmark}

\definecolor{thinkcolor}{RGB}{0,102,204}      % blue
\definecolor{searchcolor}{RGB}{102,178,255}   % light blue
\definecolor{infocolor}{RGB}{230,120,20}      % orange
\definecolor{answercolor}{RGB}{180,0,120}     % magenta

\usepackage{tikz}
\usepackage{xcolor}
\usepackage[T1]{fontenc}
\usetikzlibrary{positioning, fit, backgrounds, shapes.geometric, arrows.meta, calc}
\usepackage{float}

\definecolor{purpleFill}{HTML}{EEEDFE}
\definecolor{purpleStroke}{HTML}{534AB7}
\definecolor{purpleText}{HTML}{3C3489}
\definecolor{tealFill}{HTML}{E1F5EE}
\definecolor{tealStroke}{HTML}{0F6E56}
\definecolor{tealText}{HTML}{085041}
\definecolor{greenFill}{HTML}{EAF3DE}
\definecolor{greenStroke}{HTML}{3B6D11}
\definecolor{greenDot}{HTML}{639922}
\definecolor{greenText}{HTML}{27500A}
\definecolor{grayFill}{HTML}{F1EFE8}
\definecolor{grayStroke}{HTML}{5F5E5A}
\definecolor{grayText}{HTML}{2C2C2A}
\definecolor{blueFill}{HTML}{E8EEF1}
\definecolor{blueStroke}{HTML}{3D5A6C}
\definecolor{blueText}{HTML}{1A2B34}
\definecolor{docFill}{HTML}{F1EFE8}
\definecolor{docStroke}{HTML}{B4B2A9}
\definecolor{docText}{HTML}{444441}
\definecolor{docBody}{HTML}{5F5E5A}
\definecolor{laneStroke}{HTML}{5DCAA5}
\definecolor{lanePurple}{HTML}{AFA9EC}

\allowdisplaybreaks

\title{SubSearch: Intermediate Rewards for Unsupervised Guided Reasoning in Complex Retrieval}

\author{Roxana Petcu, Evangelos Kanoulas \& Maarten de Rijke \\
IRLab, University of Amsterdam\\
\texttt{\{r.m.petcu,e.kanoulas,m.derijke\}@uva.nl}}
\begin{document}

\ifcolmsubmission
\linenumbers
\fi

\maketitle

\begin{abstract}

Large language models (LLMs) are probabilistic in nature and perform more reliably when augmented with external information.
As complex queries often require multi-step reasoning over the retrieved information, with no clear or predetermined reasoning path, 
they remain challenging. 
Recent approaches train models using reinforcement learning on the model's outcome, showing promise in improving how models handle complex information.
We introduce SubSearch, a specialized framework that shifts from outcome-only supervision to intermediate reward signals that incentivize planning high-quality reasoning. 
Unlike previous work on process reward modeling, which focuses on training a separate reward model with annotated trajectories by either human annotators or large LLM judges, SubSearch directly optimizes the generator using \emph{intrinsic} process rewards, which we define as internally-derived rewards, eliminating the need for external supervision, and moving towards autonomous information-intensive reasoning.
Experiments on seven benchmarks show that rewarding intermediate reasoning steps with intrinsic rewards leads to more robust reasoning traces in both QA and multi-hop QA datasets over using only outcome rewards. 
SubSearch can help in building reasoning traces that allow agents to better integrate search engines for complex query answering, while offering a data-efficient alternative to supervised process modeling.\footnote{Code can be found at: % \url{https://anonymous.4open.science/r/SubSearch-B4F7/README.md}}
\url{https://github.com/RoxanaPetcu/SubSearch}}
\end{abstract}

\section{Introduction}

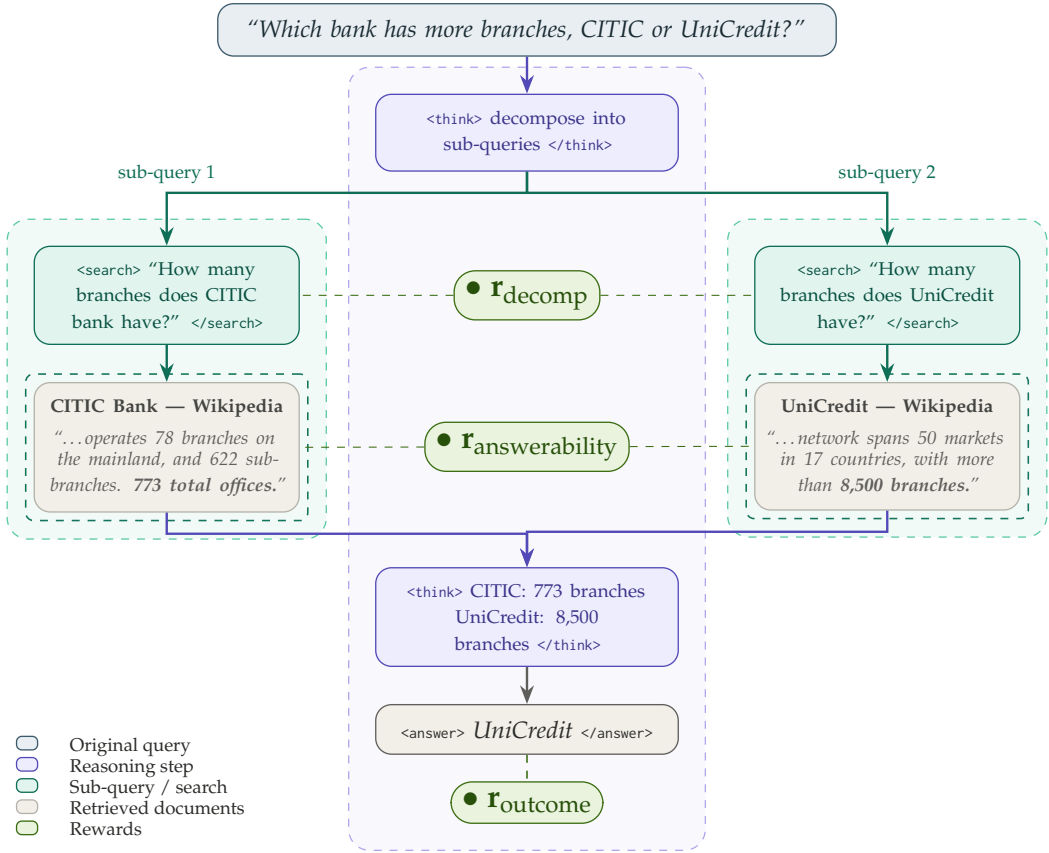
\begin{figure}[!htb]
  \centering
  %% ── TikZ styles ─────────────────────────────────────────────────
\tikzset{
  %% generic rounded box
  rbox/.style={
    draw, rounded corners=6pt, inner sep=6pt,
    text width=#1, align=center, font=\small
  },
  %% reasoning / think boxes  (purple)
  thinkbox/.style={
    rbox=#1,
    fill=purpleFill, draw=purpleStroke, line width=0.5pt,
    text=purpleText
  },
  %% sub-query / search boxes  (teal)
  searchbox/.style={
    rbox=#1,
    fill=tealFill, draw=tealStroke, line width=0.5pt,
    text=tealText
  },
  %% neutral query / answer boxes  (gray)
  graybox/.style={
    rbox=#1,
    fill=grayFill, draw=grayStroke, line width=0.5pt,
    text=grayText
  },
  bluebox/.style={
    rbox=#1,
    fill=blueFill, draw=blueStroke, line width=0.5pt,
    text=blueText
  },
  %% document snippet card
  docbox/.style={
    rbox=#1,
    fill=docFill, draw=docStroke, line width=0.5pt,
    text=docBody
  },
  %% reward pill
  rpill/.style={
    draw=greenStroke, fill=greenFill, line width=0.5pt,
    rounded corners=8pt, inner sep=4pt,
    font=\scriptsize\sffamily, text=greenText
  },
  %% dashed highlight border inside doc card
  highlight/.style={
    draw=tealStroke, dashed, line width=0.6pt, rounded corners=3pt
  },
  %% flow arrow
  arr/.style={
    -Stealth, line width=0.9pt
  },
  %% dashed lane border
  lane/.style={
    draw=#1, dashed, line width=0.5pt, rounded corners=8pt,
    fill=#1, fill opacity=0.08
  },
}

\begin{tikzpicture}[node distance=0pt]

%% ════════════════════════════════════════════════════════════════
%% ROW 1 — Query
%% ════════════════════════════════════════════════════════════════
\node[bluebox=7.8cm] (query) {%
  % \textbf{Query}\\[2pt]
  \textit{``Which bank has more branches, CITIC or UniCredit?''}%
};

%% ════════════════════════════════════════════════════════════════
%% ROW 2 — <think> decompose
%% ════════════════════════════════════════════════════════════════
\node[thinkbox=3.6cm, below=14pt of query] (think1) {%
  {\fontsize{6}{7}\selectfont \texttt{<think>}}
  {\scriptsize decompose into sub-queries}
  {\fontsize{6}{7}\selectfont \texttt{</think>}}%
};

\draw[arr, purpleStroke] (query.south) -- (think1.north);

%% ════════════════════════════════════════════════════════════════
%% ROW 3 — Sub-queries  (fan-out)
%% ════════════════════════════════════════════════════════════════
\node[searchbox=3.1cm, below left=28pt and 1cm of think1] (sq1) {%
  {\fontsize{6}{7}\selectfont \texttt{<search>}}
  {\scriptsize``How many branches does CITIC bank have?''}
  {\fontsize{6}{7}\selectfont \texttt{</search>}}\
};

\node[searchbox=3.1cm, below right=28pt and 1cm of think1] (sq2) {%
  {\fontsize{6}{7}\selectfont \texttt{<search>}}
  {\scriptsize``How many branches does UniCredit have?''}
  {\fontsize{6}{7}\selectfont \texttt{</search>}}\
};

%% fan-out arrows  (L-shaped: down then diagonal)
\draw[arr, tealStroke]
  (think1.south) -- ++(0,-8pt) -| (sq1.north);
\draw[arr, tealStroke]
  (think1.south) -- ++(0,-8pt) -| (sq2.north);

%% r_decomp pill — centred between sq1 and sq2, at their vertical midpoint
% \node[rpill, below=6pt of think1] (rdecomp)
%   {\(\bullet\)~\(\mathbf{r}_{\text{decomp}}\)};

\node[rpill, font=\large] (rdecomp) at ($(sq1.east)!0.5!(sq2.west)$)
  {\(\bullet\)~\(\mathbf{r}_{\text{decomp}}\)};

\draw[greenStroke, dashed, line width=0.4pt] (rdecomp) -- (sq1.east);
\draw[greenStroke, dashed, line width=0.4pt] (rdecomp) -- (sq2.west);

%% ════════════════════════════════════════════════════════════════
%% ROW 4 — Retrieved document cards
%% ════════════════════════════════════════════════════════════════
\node[docbox=3.1cm, below=14pt of sq1] (doc1) {%
  \textcolor{docText}{\textbf{\scriptsize CITIC Bank --- Wikipedia}}\\[4pt]
  \scriptsize\itshape
  ``\ldots operates 78 branches on\\
  the mainland, and 622 sub-\\
  branches. \textbf{773 total offices.}''%
};

\node[docbox=3.1cm, below=14pt of sq2] (doc2) {%
  \textcolor{docText}{\textbf{\scriptsize UniCredit --- Wikipedia}}\\[4pt]
  \scriptsize\itshape
  ``\ldots network spans 50 markets\\
  in 17 countries, with more\\
  than \textbf{8,500 branches.}''%
};

\draw[arr, tealStroke] (sq1.south) -- (doc1.north);
\draw[arr, tealStroke] (sq2.south) -- (doc2.north);

%% dashed highlight borders around key facts
\begin{scope}[on background layer]
  \node[highlight, fit=(doc1), inner sep=3pt] {};
  \node[highlight, fit=(doc2), inner sep=3pt] {};
\end{scope}

%% r_answerability pill
\node[rpill, font=\large] (rans) at ($(doc1.east)!0.5!(doc2.west)$)
  {\(\bullet\)~\(\mathbf{r}_{\text{answerability}}\)};

\draw[greenStroke, dashed, line width=0.4pt] (rans) -- (doc1.east);
\draw[greenStroke, dashed, line width=0.4pt] (rans) -- (doc2.west);

%% ════════════════════════════════════════════════════════════════
%% ROW 5 — <think> aggregate
%% ════════════════════════════════════════════════════════════════
\node[thinkbox=3.6cm,
      below=30pt of rans,
      yshift=-6pt] (think2) {%
  {\fontsize{6}{7}\selectfont \texttt{<think>}}
  {\scriptsize CITIC: 773 branches \\ UniCredit: 8{,}500 branches}
  {\fontsize{6}{7}\selectfont \texttt{</think>}}
};

%% merge arrows from docs back to think2  (L-shaped)
\draw[arr, purpleStroke]
  (doc1.south) -- ++(0,-8pt) -| (think2.north);
\draw[arr, purpleStroke]
  (doc2.south) -- ++(0,-8pt) -| (think2.north);

%% ════════════════════════════════════════════════════════════════
%% ROW 6 — <answer>
%% ════════════════════════════════════════════════════════════════
\node[graybox=3.6cm, below=14pt of think2] (answer) {%
  {\fontsize{6}{7}\selectfont \texttt{<answer>}}
  \textit{UniCredit}
  {\fontsize{6}{7}\selectfont \texttt{</answer>}}
};

\draw[arr, grayStroke] (think2.south) -- (answer.north);

%% ════════════════════════════════════════════════════════════════
%% ROW 7 — r_outcome pill
%% ════════════════════════════════════════════════════════════════
\node[rpill, font=\large, below=10pt of answer] (routcome)
  {\(\bullet\)~\(\mathbf{r}_{\text{outcome}}\)};

\draw[greenStroke, dashed, line width=0.6pt] (answer.south) -- (routcome.north);

%% ════════════════════════════════════════════════════════════════
%% SWIMLANE BACKGROUNDS  (drawn behind everything)
%% ════════════════════════════════════════════════════════════════
\begin{scope}[on background layer]
  %% centre reasoning lane
  \node[lane=lanePurple,
        fit=(think1)(think2)(answer)(routcome),
        inner sep=10pt,
        label={[font=\scriptsize\color{purpleStroke}, yshift=2pt]above:reasoning trace}
       ] {};

  %% left sub-query lane
  \node[lane=laneStroke,
        fit=(sq1)(doc1),
        inner sep=10pt,
        label={[font=\scriptsize\color{tealStroke}, yshift=10pt]above:sub-query 1}
       ] {};

  %% right sub-query lane
  \node[lane=laneStroke,
        fit=(sq2)(doc2),
        inner sep=10pt,
        label={[font=\scriptsize\color{tealStroke}, yshift=10pt]above:sub-query 2}
       ] {};
\end{scope}

%% ════════════════════════════════════════════════════════════════
%% LEGEND  (Bottom Left)
%% ════════════════════════════════════════════════════════════════
\node[anchor=south west, 
      at={(current bounding box.south west)}, 
      yshift=0pt, % Adjust this to move it further down
      font=\scriptsize, 
      text=grayText] (legend) {%
  \begin{tabular}{@{}cl@{}}
    \tikz\draw[fill=blueFill,draw=blueStroke,line width=0.5pt,rounded corners=2pt] (0,0) rectangle (0.28,0.18); & Original query \\
    \tikz\draw[fill=purpleFill,draw=purpleStroke,line width=0.5pt,rounded corners=2pt] (0,0) rectangle (0.28,0.18); & Reasoning step \\
    \tikz\draw[fill=tealFill,draw=tealStroke,line width=0.5pt,rounded corners=2pt] (0,0) rectangle (0.28,0.18); & Sub-query / search \\
    \tikz\draw[fill=docFill,draw=docStroke,line width=0.5pt,rounded corners=2pt] (0,0) rectangle (0.28,0.18); & Retrieved documents \\
    \tikz\draw[fill=greenFill,draw=greenStroke,line width=0.5pt,rounded corners=2pt] (0,0) rectangle (0.28,0.18); & Rewards
  \end{tabular}%
};

\end{tikzpicture}
  \caption{SubSearch query decomposition and intermediate reward computation. The rewards are calculated at different stages of the reasoning traces, but only after the entire reasoning trace has been generated.}
  \label{fig:subsearch}
\end{figure}

Focus on model reasoning has shifted from simple question-answering (QA) \citep{liu2024chatqa} to information-intensive complex tasks, for which current large language models (LLMs) still face challenges, such as the need for external information \citep{wei2022chain}. %Addressing these challenges requires having the ability to retrieve the most up-to-date relevant information needed \citep{jin2024long, schick2023toolformer} by interacting with search engines instead of a static retrieval system.
Search agents have emerged as specialized retrieval-augmented generation (RAG) systems that, compared to traditional RAG systems that rely on static retrieval from a fixed database, treat search as a dynamic tool \citep{lewis2020retrieval, yao2022react}. %, generating queries, performing search and retrieving documents, ensuring optimal generation \citep{schick2023toolformer}. 
Dynamic search provides access to vast, up to date information, however a challenge still exists in how to reason over and aggregate retrieved information into useful knowledge \citep{huang2023towards}. Previous work has looked at how to guide the reasoning process through prompt-based approaches \citep{wu2023clipself, trivedi2023interleaving}, or supervised fine-tuning (SFT) \citep{asai2023self, schick2023toolformer}. As the possible reasoning trajectories of the LLM effectively cover an infinite search space, supervised methods cannot scale, blocking them from generalizing to multi-step reasoning for unpredictable, real-world information retrieval tasks.
% Dynamic search provides access to vast, up to date information, however models struggle with reasoning  \citep{petcu2025comprehensivetaxonomynegationnlp}, a challenge still existing in how to reason over and aggregate retrieved information \citep{huang2023towards}. Previous work has looked at how to guide the reasoning process through prompt-based approaches \citep{wu2023clipself, trivedi2023interleaving}, or supervised fine-tuning (SFT) \citep{asai2023self, schick2023toolformer}. As the possible reasoning trajectories of the LLM effectively cover an infinite search space, supervised methods cannot scale, blocking them from generalizing to multi-step reasoning for unpredictable, real-world information retrieval tasks.

A shift towards optimization with reinforcement learning with verifiable rewards (RLVF) has addressed the generalizability concerns of SFT \citep{jin2025search}. \citet{jin2025search} train an LLM to interleave reasoning and search by reasoning over the task, generating a search query, and calling a dynamic search tool to retrieve relevant documents. Their model then aggregates the retrieved information to evaluate its knowledge and, if insufficient, reformulates the query to apply subsequent steps until the knowledge gap dissapears. Once the aggregated information is sufficient to answer the complex query, the model generates a response that conditions a sparse, outcome-based reward used for model training.

Following this paradigm shift, multiple aspects of RLVF-based agents have been explored, e.g., tool calling \citep{ma2025otcoptimaltoolcalls, wu2025masksearchuniversalpretrainingframework}, formatting \citep{zhao2025rsearchempoweringllmreasoning, wu2025mmsearchr1incentivizinglmmssearch}, evidence generation instead of retrieval \citep{sun2025zerosearchincentivizesearchcapability}, or parallel query decomposition \citep{zhao2025parallelsearchtrainllmsdecompose}. Most prior work keeps the outcome-based rewards fixed. However, reliance on outcome-only rewards enables reward hacking  \citep{gao2022scalinglawsrewardmodel}, where a model can reach a correct conclusion through flawed intermediate reasoning. %, limiting its ability to generalize on unseen data and tasks.

We propose SubSearch, a framework for training deep search agents using intermediate reasoning rewards that incentivize the generation and decomposition of complex reasoning traces. Unlike process reward models that rely on external supervision, we introduce \textit{intrinsic} process rewards as internally-derived signals conditioned on the generator alone. We propose: (1) a template for decomposing a complex task into subqueries used for dynamic search, (2) a policy that assigns intermediate rewards at the subquery %search 
level %of the reasoning trace
, and (3) a comparison of aggregation methods for constructing a stable and informative signal.

% (1) How can we model a query tree decomposition for search? (2) How can we model subquery-level intermediary rewards? (3) How can we introduce intermediary rewards to the final sparse reward, use it for RLVF, and how does that affect the final answer generation?

% NOTE: keep in mind other previous work looked into (1) parallel decomposition: ParallelSearch, TreeSearch, GlobalRAG, but without complex rewards, and others looked into (2) composed rewards: InForage, O2-Searcher but not intermediate, and (3) R-Search looked into an intermediate reward but one that still is only applied at the level of the generate answer. So far there has not been a study that looks into intermediate rewards signaling different components of the reasoning pipeline and doing so in a parallel decomposition.
\section{Related Work}

\subsection{Reinforcement Learning}
In reinforcement learning (RL) an agent adapts while learning from an environment by taking actions and receiving feedback, reinforcing the agent's beliefs of the environment. RL has been incorporated into LLMs through human feedback (RLHF) \citep{kaufmann2024surveyreinforcementlearninghuman}, and through RLVF. %where the feedback provided to the agent is in the form of human preferences. 
The model update based on feedback is often made through algorithms such as proximal policy optimization (PPO) \citep{schulman2017proximalpolicyoptimizationalgorithms}, direct preference optimization (DPO) \citep{rafailov2024directpreferenceoptimizationlanguage}, %SimPO \citep{meng2024simposimplepreferenceoptimization}, ORPO \citep{hong2024orpomonolithicpreferenceoptimization}, 
or group relative policy optimization (GRPO) \citep{deepseekai2025deepseekr1incentivizingreasoningcapability}. In contrast, supervised fine-tuning (SFT) trains the model using annotations, requiring extensive resources. While often reaching better effectiveness, it can cause models to memorize solution paths, bringing generalizability limitations and evidence that reasoning is not achieved \citep{chu2025sftmemorizesrlgeneralizes}.

\subsection{Deep Search Agents}
LLMs are powerful reasoners~\citep{grattafiori2024llama3herdmodels,openai2024gpt4technicalreport,geminiteam2025geminifamilyhighlycapable}, however, their performance is conditioned on domain-specific knowledge \citep{mallen2023trustlanguagemodelsinvestigating}, which is often insufficient. RAG addresses the knowledge gap by incorporating external information as context to the model. The main challenge is knowing how to reason over these external sources and aggregate them to form an answer \citep{jin2025search}. Recent frameworks such as Search-R1 employ reinforcement learning to develop specialized search policies, where the model iteratively reasons about its knowledge and refines its search trajectory. This paradigm offers an effective solution to reasoning over relevant documents.

\header{Environment and tool optimization} Several frameworks cater for the interaction between agents and dynamic search systems. DeepResearcher \citep{zheng2025deepresearcher} treats search as an interactive engine using web search interactions instead of retrieving from a fixed corpus. OTC \citep{ma2025otcoptimaltoolcalls} and  MaskSearch \citep{wu2025masksearchuniversalpretrainingframework} optimize for calling the search engine. ZeroSearch \citep{sun2025zerosearchincentivizesearchcapability} eliminates the search engine entirely, and trains the model to generate documents instead of retrieving.

\header{Knowledge needs} Another challenge addressed by previous work is deciding on when to search. IKEA \citep{huang2025reinforcedinternalexternalknowledgesynergistic} optimizes the model to search only when crucial information does not already exist in its parametric knowledge. R-Search \citep{zhao2025rsearchempoweringllmreasoning} does not control when the search engine is called. It can be called at any generated token. InForage \citep{qian2025scentknowledgeoptimizingsearchenhanced} and O\(^2\)-Searcher \citep{mei2025o2searchersearchingbasedagentmodel} apply specialized SFT using human-guided search reasoning datasets with annotated reasoning trajectories.

\header{Architecture} TreeSearch \citep{koh2026treesearchlanguagemodel} has a different schema for GRPO, modeling it as a tree search structure, where each tree node represents a complete agent interaction step, effectively designing an orchestration system. ParallelSearch \citep{zhao2025parallelsearchtrainllmsdecompose} trains a model to decompose the query and apply search for each. Similarly, GlobalRAG \citep{luo2026globalragenhancingglobalreasoning} decomposes questions into subgoals.

\subsection{Credit Assignment in Deep Search Agents}

An important aspect of training deep search agents is credit assignment, i.e., assigning rewards to reasoning steps that contribute to the final answer generation. To mitigate reward hacking from outcome-only supervision, process reward models (PRMs) \citep{lightman2023letsverifystepstep} shifted to intermediate signals to guide the model through the reasoning trace. Frameworks such as RAG-Gym \citep{xiong2025raggymsystematicoptimizationlanguage} and ReasonRAG \citep{zhang2025processvsoutcomereward} explicitly train a reward model with human feedback or LLM judges to become better selectors of generated reasoning traces. Unlike previous work on PRMs, SubSearch directly optimizes the generator using intrinsic process rewards, where a process reward is intrinsic if it is derived only from the model's state, such as semantic coverage, rather than from an external annotator, thus eliminating the need for additional resources and moving towards autonomous information-intensive reasoning.

\section{SubSearch}

In this section we introduce SubSearch, a process-based deep search agent with intermediate rewards that assess and quantify the quality of reasoning decomposition and query rewrites without using manually annotated reasoning trajectories for SFT. SubSearch decomposes a complex information need into subqueries, and interacts with the search environment to retrieve relevant documents for each. 
%The optimal performance of SubSearch comes from integrating intermediary rewards during training, where each subquery is evaluated and has a reward attributed. In this section, we will cover the theoretical background on deep search agents needed for this paper, the training template, intermediate rewards signals, and training paradigm.

\subsection{Preliminaries}
\header{Dynamic search} %Deep Research looks into optimizing RAG pipelines from different angles to achieve better answer generation given complex queries. One core aspect of advanced RAG pipelines is to guide the model to reason and to search, enabling the model to create its own reasoning trace, and retrieve evidence at each step. In this process, the model can determine its next action based on its current observation of the search environment. 
Deep search agents produce a reasoning trajectory signaled by tokens that trigger specific actions: the internal reasoning is wrapped within $\langle\texttt{think}\rangle$ and $\langle\texttt{/think}\rangle$, search queries are generated within $\langle\texttt{search}\rangle$ and $\langle\texttt{/search}\rangle$, retrieved documents within $\langle\texttt{information}\rangle$ and $\langle\texttt{/information}\rangle$, while the final generation is between $\langle\texttt{answer}\rangle$ and $\langle\texttt{/answer}\rangle$. The iterative process ends once there is sufficient information to generate an answer. The trace follows a multi-turn reasoning-search loop:
\begin{equation}
    (t_0, s_0, c_0, ..., t_{n-1}, s_{n-1}, c_{n-1}, t_n, a),
\end{equation}
where \(t\) is the thinking process, \(s\) the search action, \(c\) the retrieved context, and \(a\) the answer.

\header{Reinforcement learning} 
A commonly used formulation of the RL objective using a search engine \(\mathcal{R}\) is as follows:
\begin{equation}
\max_{\pi_\theta} \mathbb{E}_{x \sim \mathcal{D}, a \sim \pi_\theta (\cdot|x; \mathcal{R})} \left[ r_\phi(x, a) \right] - \beta D_{\text{KL}} \left[ \pi_\theta(a | x; \mathcal{R}) \, \Vert \, \pi_{\text{ref}}(a | x; \mathcal{R}) \right],
\end{equation}
where \(\mathcal{R}\) denotes the search engine, \(x\) is the input query sampled from the data distribution \(\mathcal{D}\), \(a\) represents the output sequence, \(\pi_{\theta}\) denotes the policy LLM, \(\pi_\text{ref}\) is the reference LLM, and \(\pi_{\phi}\) denotes the reward function.

Search agents can be trained, among others, with group relative policy optimization (GRPO) \citep{deepseekai2025deepseekr1incentivizingreasoningcapability} as seen in Eq.~\ref{eq:grpo}:
\begin{equation}
\begin{aligned}
\mbox{}\hspace*{-2mm}
\mathcal{J}_{GRPO}(\theta) = \mathbb{E} \!\left[ q \sim P(Q), \{o_i\}_{i=1}^G \sim \pi_{\theta_{old}} \right] \frac{1}{G} \sum_{i=1}^G \frac{1}{|o_i|} \sum_{t=1}^{|o_i|} \bigg( \min \left( \frac{\pi_\theta(o_{i,t} | q, o_{i,<t})}{\pi_{\theta_{old}}(o_{i,t} | q, o_{i,<t})} \hat{A}_{i,t}, \right. \\
\quad \left. \text{clip} \left( \frac{\pi_\theta(o_{i,t} | q, o_{i,<t})}{\pi_{\theta_{old}}(o_{i,t} | q, o_{i,<t})}, 1-\epsilon, 1+\epsilon \right) \hat{A}_{i,t} \right) - \beta D_{KL}(\pi_\theta \| \pi_\text{ref}) \bigg),
\end{aligned}
\label{eq:grpo}
\end{equation}

% \begin{equation}
% \mbox{}\hspace*{-1.5mm}
% \mathcal{J}_{GRPO}(\theta)\!=\! \mathbb{E}\!\left[ q \!\sim\! P(Q), \{o_i\}_{i=1}^G \!\sim\! \pi_{\theta_{old}}(O|q) \right]\!\frac{1}{G} \sum_{i=1}^G\!\left( \!\frac{1}{|o_i|} \sum_{t=1}^{|o_i|} \left( \mathcal{L}_{i,t}(\theta) \right) \!-\! \beta D_{KL}(\pi_\theta \| \pi_\mathit{ref})\!\right)
% \hspace*{-1.5mm}\mbox{}
% \label{eq:grpo}
% \end{equation}

% \begin{equation}\label{eq:grpo_loss}
% \mathcal{L}_{i,t}(\theta) = \min \left( \frac{\pi_\theta(o_{i,t} | q, o_{i,<t})}{\pi_{\theta_{old}}(o_{i,t} | q, o_{i,<t})} \hat{A}_{i,t}, \text{clip} \left( \frac{\pi_\theta(o_{i,t} | q, o_{i,<t})}{\pi_{\theta_{old}}(o_{i,t} | q, o_{i,<t})}, 1-\epsilon, 1+\epsilon \right) \hat{A}_{i,t} \right)
% \end{equation}

where \(\epsilon\) and \(\beta\) are hyperparameters, and \(\hat{A}_{i,t}\) represents the advantage calculated based on the relative rewards of all outputs generated within each group.

\subsection{Training Template}

We apply a multi-turn interaction template that guides the policy model through iterative reasoning and information retrieval until a final answer is reached. Previous work has showed that the decomposition of complex queries leads to better retrieval, and that retrieval conditioned relevance signals can be used for estimating subquery utility \citep{petcu2025querydecompositionragbalancing}. We incentivize the model to decompose the query into subqueries at each step if needed, allowing the reasoning trace to perform both sequential and parallel decompositions, depending on the reasoning type found in the initial query. Table \ref{tab:training_template} illustrates our template.

\begin{table}[htbp]
\centering
\small
\begin{tabular}{p{0.95\linewidth}} 
    \toprule
    Answer the given question. You must conduct reasoning inside \textcolor{thinkcolor}{\textbf{\textless think\textgreater}} and \textcolor{thinkcolor}{\textbf{\textless/think\textgreater}} first every time you get new information. After reasoning, if you find you lack some knowledge, you can call a search engine by \textcolor{searchcolor}{\textbf{\textless search\textgreater}} \textit{query} \textcolor{searchcolor}{\textbf{\textless/search\textgreater}}, and it will return the top searched results between \textcolor{infocolor}{\textbf{\textless information\textgreater}} and \textcolor{infocolor}{\textbf{\textless/information\textgreater}}. \\
    \addlinespace[0.8em]
    If the original query is complex or involves multiple parts, you are encouraged to \textbf{decompose} it into at most 3 smaller sub-questions, separated by \texttt{\#\#}. For example: \textcolor{searchcolor}{\textbf{\textless search\textgreater}} sub-question 1 \#\# sub-question 2 \textcolor{searchcolor}{\textbf{\textless/search\textgreater}} and it will return the top searched results between \textcolor{infocolor}{\textbf{\textless information\textgreater}} documents sub-question 1 \#\# documents sub-question 2 \textcolor{infocolor}{\textbf{\textless/information\textgreater}}. \\
    \addlinespace[0.8em]
    You can search as many times as you want. Only \textbf{decompose} when the question has multiple independent parts (e.g., different entities, aspects, or comparisons). Do not decompose questions that do not need it. \\
    \addlinespace[0.8em]
    If you find no further external knowledge needed, you can directly provide the answer inside \textcolor{answercolor}{\textbf{\textless answer\textgreater}} and \textcolor{answercolor}{\textbf{\textless/answer\textgreater}} without detailed illustrations. For example, \textcolor{answercolor}{\textbf{\textless answer\textgreater}} Beijing \textcolor{answercolor}{\textbf{\textless/answer\textgreater}}. \\
    \addlinespace[0.5em]
    \textbf{Question:} \{question\} \\ 
    \bottomrule
\end{tabular}
\caption{Prompt template for SubSearch. The question is appended during training and inference.}
\label{tab:training_template}
\end{table}

\subsection{Intermediate Reward Modeling}

SubSearch integrates intermediate rewards conditioned on answerability \citep{rajpurkar2018knowdontknowunanswerable} calculated at each subquery, and decomposition \citep{fu2021decomposingcomplexquestionsmakes} calculated for each (sub)query that is split. For further details on the notation, see Figure \ref{fig:query_decomp_and_rewards} (Appendix \ref{appendix:decomposition_and_rewards}).

\header{Answer reward} We follow the classic outcome-based signal for answer generation. We calculate the exact string matching (EM) between the generated and the gold answer, a rule-based binary metric:
\begin{equation}
    \mathcal{R}_\text{answer}(a) = EM(a, a_\text{gold}).
\end{equation}

\header{Subquery answerability} The \textit{answerability} of a subquery reflects how well the search engine addresses that subquery through its retrieved ranked list of documents. It is the cosine similarity between the subquery embedding and the top-$k$ retrieved document embeddings as measured by an encoder, serving as a continuous proxy signal for information coverage:
\begin{equation}
\begin{aligned}
\mathcal{R}_{\text{answerability}}^{(l,i)}(x_{l,i}, D_{l,i}) &= \frac{1}{k} \sum_{d_{i,j} \in \text{top-$k$}(D_{l,i})} \text{sim}\left(\phi(x_{l,i}), \phi(d_{i,j})\right),
\end{aligned}
\label{eq:reward_answerability}
\end{equation}
where $x_{l,i}$ is the subquery at decomposition level $l \in [1, L]$ and index \(i\), $D_i$ represents the retrieved ranked list of documents, $\phi{(\cdot)}$ is an embedding model, and \(sim(\cdot)\) calculates cosine similarity as a search similarity score.

\header{(Sub)query decomposition} The decomposition reward %at any level $l \in [1, L]$, where we consider the levels to increase from the root to the last decompositions, 
is a weighted combination of two distinct objectives. First, semantic coverage $r_{\text{coverage}}^{(l)}$ ensures the aggregated subqueries at level $l$ mantain the same information as the parent query at level $l-1$, which is calculated as the cosine similarity between the average embeddings at level $l$ with the parent query embedding at $l-1$, preventing ``query drift.'' Second, in-group splitability $r_{\text{split}}^{(l)}$ maximizes the product of a subquery’s relevance to its parent and its uniqueness relative to its siblings at the same level $l$. This dual-constraint approach ensures that each decomposition step produces subqueries that are collectively exhaustive but mutually exclusive in their information requirements:
%
% \begin{equation}
% \begin{aligned}
% r_{\text{coverage}}(x_{i-1}, \{x_i\}) &= \cos\left(x_{i-1},\ \frac{1}{n}\sum_{i=1}^{n} x_i\right) \\
% \quad r_{\text{splitability}}(x, \{x_i\}) &= \frac{1}{n}\sum_{i=1}^{n} \text{sim}(x, \{x_i\}) \cdot \left(1 - \frac{1}{n-1}\sum_{j \neq i} \text{sim}(\{x_i\}, \{x_j\})\right) \\
% \qquad r_{\text{decomp}}(x_{i-1}, \{x_i\}) &= \alpha \cdot r_{\text{coverage}}(x_{i-1}, \{x_i\}) + \beta \cdot r_{\text{split}}(x_{i-1}, \{x_i\})
% \end{aligned}
% \label{eq:reward_structure}
% \end{equation}
% where $\{x_i\}$ represents the subqueries decomposed of the previously generated (sub)query $\{x_{i-1}\}$, \(\alpha=0.5\), and \(\beta=0.5\) are set coefficients.
%
\begin{equation}
\begin{aligned}
r_{\text{coverage}}(x_{l-1}, \{x_l\}_i^n) &= \text{sim}\left(\phi(x_{l-1}),\ \frac{1}{n}\sum_{i=1}^{n} \phi(x_{l,i})\right) \\
\quad r_{\text{split}}(x_{l-1}, \{x_l\}_i^n) &= \frac{1}{n}\sum_{i=1}^{n} \left[ \text{sim}(\phi(x_{l-1}), \phi(x_{l,i})) \cdot \left(1 - \frac{1}{n-1}\sum_{j \neq i}^n(\phi(x_{l,i}), \phi(x_{l,j}))\right) \right] \\
\qquad \mathcal{R}_{\text{decomposition}}^{(l)} &= \alpha \cdot r_{\text{coverage}} + \beta \cdot r_{\text{split}},
\end{aligned}
\label{eq:reward_decomposition}
\end{equation}
where $\{x_l\}_1^n$ represents the subqueries decomposed from the previously generated (sub)query $\{x_{l-1}\}$, \(\phi\) is the embedding model, \(\alpha\), and \(\beta\) are set coefficients.

\header{Format reward}
In addition to the previously described rewards, we introduce a format reward to stabilize training:
%
% %
% \begin{equation}
% r_{\text{form}} =
% \begin{cases}
% -\lambda_\mathit{structure} & \text{if } r \neq 0 \wedge f_\mathit{format} = \mathit{False} \\
% \lambda_\mathit{retrieval} & \text{if } r \neq 0 \wedge f_\mathit{format} = \mathit{True} \wedge f_{retrieval} = True \\
% \lambda_\mathit{format} & \text{if } r \neq 0 \wedge f_\mathit{format} = \mathit{True} \wedge f_\mathit{retrieval} = \mathit{False} \\
% \lambda_\mathit{retrieval} & \text{if } r = 0 \wedge f_\mathit{format} = \mathit{True} \wedge f_\mathit{retrieval} = \mathit{True} \\
% \lambda_\mathit{structure} & \text{if } r = 0 \wedge f_\mathit{format} = \mathit{True}  \\
% 0 & \text{if } r = 0 \wedge f_\mathit{format} = \mathit{False},\\
% \end{cases}
% \end{equation}
% %
%
\begin{equation}
r_\text{format} = \begin{cases}
0 & \text{if } f_\text{format} = \text{False} \land f_\text{retrieval} = \text{False} \\
\lambda_\text{structure} & \text{if } f_\text{format} = \text{True} \land f_\text{retrieval} = \text{False} \\
\lambda_\text{structure} + \lambda_\text{retrieval} & \text{if } f_\text{format} = \text{True} \land f_\text{retrieval} = \text{True}, \\
\end{cases}
\end{equation}
where \(\lambda_\text{format}\) and \(\lambda_\text{retrieval}\) are set values.

% \begin{equation}
% r_{\phi}(x, y) = 
% \begin{cases} 
% 1 & \text{if } a_{pred} = a_{gold} \wedge f_{format}(y) = \text{True}, \\
% 1 - \lambda_f & \text{if } a_{pred} = a_{gold} \wedge f_{format}(y) = \text{False}, \\
% \lambda_f & \text{if } a_{pred} \neq a_{gold} \wedge f_{format}(y) = \text{True}, \\
% 0 & \text{if } a_{pred} \neq a_{gold} \wedge f_{format}(y) = \text{False}.
% \end{cases}
% \end{equation}

\header{Aggregation} We aggregate the intermediate continuous rewards with the final sparse reward using adaptive residual reward aggregation:
%
% \begin{equation}
%     r = r_{ans} + \beta * (1 -r_{ans}) * \Bigl[ \frac{1}{m}\sum r_{\text{inter}}(x, x_i) \Bigr] + r_{form}
% \end{equation}
%
% \begin{equation}
% \begin{aligned}
% r = \mathcal{R}_{\text{answer}} + \beta (1 - \mathcal{R}_{\text{ans}}) \cdot \frac{1}{2} \left[ \frac{1}{M} \sum_{j=1}^{M} \mathcal{R}_{\text{answerability}}^{(j)} + \frac{1}{L} \sum_{l=1}^{L} \mathcal{R}_{\text{decomp}}^{(l)} \right] + r_{form}
% \end{aligned}
% \label{eq:adaptive_aggregation_simple}
% \end{equation}
% where \(M\) is the number of generated subqueries, and \(L\) the number of decomposed levels.
%
\begin{equation}
\begin{aligned}
r = \mathcal{R}_{\text{answer}} + \beta (1 - \mathcal{R}_{\text{answer}}) \cdot \frac{1}{2} \left[avg(\mathcal{R}_\text{answerability}) + avg(\mathcal{R}_\text{decomposition}) \right] + r_\text{form},
\end{aligned}
\label{eq:adaptive_aggregation_simple}
\end{equation}
where the answerability and decomposition rewards are averaged over the number of subqueries, and the number of decomposed queries respectively.
\section{Experiments and Results}

\subsection{Datasets and Models}
To evaluate the effectiveness of SubSearch, we adopt the evaluation setup established by Search-R1 \citep{jin2025search}, which covers seven benchmarks. These include open-domain QA tasks such as Natural Questions (NQ), TriviaQA, and PopQA, alongside multi-hop reasoning QA datasets such as HotpotQA, 2WikiMultiHopQA, MuSiQue, and Bamboogle.

\subsection{Baselines}
We evaluate SubSearch against standard inference methods, i.e., Direct Inference, CoT, and RAG, alongside state-of-the-art RL-based search agents; see Table~\ref{tab:method_comparison}. We categorize the advanced baselines based on their supervision requirements:

\begin{itemize}[leftmargin=*]
    \item Search-R1 \citep{jin2025search} and ZeroSearch \citep{sun2025zerosearchincentivizesearchcapability} optimize for global outcome rewards using GRPO. R-Search \citep{zhao2025rsearchempoweringllmreasoning} extends search by allowing token-level retrieval triggers and uses an auxiliary LLM judge (Llama 3.2 3B) to generate intermediate evidence-quality signals. R-Search was initially trained on MuSiQue and was evaluated using \(\text{top-}k=5\) retrieved documents. To make it comparable with the other methods, we reproduce this approach with the standard setup of training on a merged dataset of NQ and HotpotQA and using \(\text{top-}k=3\) retrieved documents.
    
    \item InForage \citep{qian2025scentknowledgeoptimizingsearchenhanced} and O$^2$-Searcher \citep{mei2025o2searchersearchingbasedagentmodel} rely on SFT with specialized, human-annotated datasets with reasoning trajectories. InForage further incorporates an information-gain reward that requires access to golden documents, while O$^2$-Searcher uses a diversity reward and performs explicit knowledge gap updates. Both represent a high-cost upper bound due to their dependence on expert annotations.
\end{itemize}

\subsection{Experimental Setup}

We train SubSearch by merging the NQ and HotpotQA datasets, and evaluate using EM. We use Qwen3.2-3B-base and -instruct as s to our model. We train the base model for 600 steps and the instruct model for 200, as the instruct model tends to collapse earlier. For the answerability reward we use \(\text{top-}k=3\) documents due to efficiency constraints, for the decomposition rewards we set coefficients \(\alpha=0.5\) and \(\beta=0.5\), and for the format reward we use \(\lambda_\text{structure} = \lambda_\text{retrieval} = 0.1\). Further details can be found in Appendix \ref{appendix:experimental_setup}.

\begin{table}[t]
\centering
\small
\setlength{\tabcolsep}{3pt}
\renewcommand{\arraystretch}{1.1}
\begin{tabular}{lcllclcc}
\toprule
\multirow{2}{*}{\textbf{}} &
\multirow{2}{*}{\textbf{}} &
\multirow{2}{*}{\textbf{}} &
\multicolumn{3}{c}{\textbf{RL}} &
\multirow{2}{*}{\textbf{}} &
\multirow{2}{*}{\textbf{}} \\
\cmidrule(lr){4-6}
\textbf{Method} & \textbf{Evidence} & \textbf{SFT} & \textbf{Rewards} & \textbf{Type} & \textbf{Training data} & \textbf{Decomp.} & \textbf{Interm.} \\
\midrule
Direct Inference      & --   & --   & --  & --           & --              & --   & --       \\
CoT                   & --   & --   & --  & --           & --              & --   & --       \\
SFT                   & --   & \cmark  & --  & --           & --              & --   & --       \\
\midrule
RAG                   & \cmark   & --   & --  & --           & --              & --   & --       \\
Search-o1             & \cmark   & --   & --  & --           & --              & --   & --       \\
R1-base               & --       & --   & EM  & global       & NQ+HotpotQA    & --   & --       \\
Search-R1             & \cmark   & --   & EM  & global       & NQ+HotpotQA     & --   & -- \\
ZeroSearch            & --       & --   & EM  & global       & NQ+HotpotQA & -- & -- \\
StepSearch            & \cmark   & --   & EM  & global       & MuSiQue & \cmark & -- \\
R-Search              & \cmark   & --   & EM+evidence q.  & global     & 2wiki    & --   & -- \\
\midrule
InForage              & \cmark   & \cmark   & EM+gain+eff. & global & NQ+HotpotQA    & -- & --\\
O$^2$-Searcher        & \cmark   & \cmark   & EM+fact+div. & global & NQ+HotpotQA    & -- & --\\
\midrule
\textbf{SubSearch}    & \cmark   & --   & EM+ans+dec. & \textbf{1g+2i} & NQ+HotpotQA & \cmark & \cmark \\
\bottomrule
\end{tabular}
\caption{Comparison of search agent methods. \cmark/--: present/absent. \textit{Evidence}: retrieved external documents. \textit{Decomp.}: query decomposition. \textit{Interm.}: subquery-level intermediate reward. \textbf{1g+2i}: 1 global + 2 intermediate.}
\label{tab:method_comparison}
\end{table}

\subsection{Performance}

Table \ref{tab:main_results} presents the main results of SubSearch compared to baselines across both general QA and multi-hop QA benchmarks. Overall, SubSearch consistently improves over the other RL-based search agents on both simple QA and multi-hop QA reasoning datasets. Compared to Search-R1-base, SubSearch achieves significant gains on complex benchmarks such as HotpotQA (+6.5 EM), 2WikiMultiHopQA (+7.7 EM), MuSiQue (+3.5 EM), and Bamboogle (+13.5), highlighting the effectiveness of modeling intermediate search behavior beyond final-answer supervision, while on other datasets we achieve improvements such as on NQ (+4.2 EM), TriviaQA (+1.4 EM), and PopQA (+2.2 EM).

We observe that query decomposition with EM rewards already achieves improvements over the baselines as seen in Table \ref{tab:main_results}, but incorporating intermediate reward signals further boosts performance, especially on datasets with inherently compositional structure. For example for Bamboogle, adding intermediate rewards leads to substantial gains over decomposition-only variants, suggesting that explicitly rewarding subquery quality helps the model better navigate complex reasoning–retrieval interactions.

%In contrast, improvements on general QA datasets such as NQ and TriviaQA are not significant, and in some cases plateau, indicating that decomposition and intermediate rewards are most beneficial when the task requires aggregating information over multiple documents. This aligns with our hypothesis that intermediate reward modeling is particularly effective for complex, multi-hop information needs, while offering limited benefit for factoid-style queries.

Finally, without relying on annotated reasoning trajectories or additional training data, SubSearch achieves competitive performance with methods such as InForage and O2-Searcher (see SubSearch vs. SFT+RL methods), demonstrating that intermediate reward design alone can provide a strong and scalable training signal for search agents.

\begin{table*}[ht]
\centering
\small
\setlength{\tabcolsep}{1mm}
\resizebox{\textwidth}{!}{
\begin{tabular}{llccccccc}
\toprule
\multirow{2}{*}{\textbf{Methods}} & \multirow{2}{*}{\textbf{Alg.}} & \multicolumn{3}{c}{\textbf{General QA}} & \multicolumn{4}{c}{\textbf{Multi-Hop QA}} \\
\cmidrule(lr){3-5} \cmidrule(lr){6-9}
& & \textbf{NQ$^\dagger$} & \textbf{TriviaQA$^\ast$} & \textbf{PopQA$^\ast$} & \textbf{HotpotQA$^\dagger$} & \textbf{2wiki$^\ast$} & \textbf{MuSiQue$^\ast$} & \textbf{Bamboogle$^\ast$} \\
\midrule

\textit{Baselines} & & & & & & & & \\
Direct Inference & -- & 0.106 & 0.288 & 0.108 & 0.149 & 0.244 & 0.020 & 0.024 \\
CoT              & -- & 0.023 & 0.032 & 0.005 & 0.021 & 0.021 & 0.002 & 0.000 \\
SFT              & -- & 0.249 & 0.292 & 0.104 & 0.186 & \textbf{0.248} & 0.044 & 0.112 \\
RAG              & -- & \textbf{0.348} & \textbf{0.544} & \textbf{0.387} & \textbf{0.255} & 0.226 & 0.047 & 0.080 \\
Search-o1        & -- & 0.238 & 0.472 & 0.262 & 0.221 & 0.218 & \textbf{0.054} & \textbf{0.320} \\
\midrule
\textit{RL (SFT-Free)} & & & & & & & & \\
R1-base          & PPO   & 0.226 & 0.455 & 0.173 & 0.201 & 0.268 & 0.055 & 0.224 \\
StepSearch-base   & PPO   & --    & --    & --    & 0.329 & 0.339 & \textbf{0.181} & 0.328 \\
StepSearch-instruct & PPO   & --   & --    & --    & 0.345 & 0.320 & 0.174 & \textbf{0.344} \\
ZeroSearch-instruct & REINF. & 0.414 & 0.574 & \textbf{0.448} & 0.274 & 0.300 & 0.098 & 0.111 \\
Search-R1-base   & GRPO  & 0.421 & 0.583 & 0.413 & 0.297 & 0.274 & 0.066 & 0.128 \\
Search-R1-instruct  & GRPO  & 0.397 & 0.565 & 0.391 & 0.331 & 0.310 & 0.124 & 0.232 \\
ZeroSearch-base  & GRPO  & 0.394 & 0.558 & 0.390 & 0.298 & 0.322 & 0.124 & 0.236 \\
R-Search-instruct (r.)   & GRPO  & 0.319 & 0.541 & 0.365 & 0.307 & 0.320 & 0.119 & 0.208 \\
% \addlinespace[0.3em]
Query Decomp-base (Ours) & GRPO & 0.401 & 0.560 & 0.404 & 0.302 & 0.334 & 0.099 & 0.156 \\
\rowcolor{yellow!15}
SubSearch-base (Ours)    & GRPO & \textbf{0.463} & \textbf{0.597} & 0.435 & \textbf{0.362} & \textbf{0.351} & 0.101 & 0.263 \\

\midrule
\textit{SFT + RL (Upper Bound)} & & & & & & & & \\
InForage-instruct    & PPO  & 0.421 & \textbf{0.597} & \textbf{0.452} & \textbf{0.409} & \textbf{0.428} & \textbf{0.172} & \textbf{0.360} \\
O2-Searcher-instruct & GRPO & \textbf{0.444} & \textbf{0.597} & 0.429 & 0.388 & 0.374 & 0.160 & 0.344 \\
\bottomrule
\end{tabular}%
}
\caption{Main results. Bold indicates the best performance within each supervision category. $\dagger$/$\ast$ represent in-domain/out-domain datasets. \textit{REINF.} refers to the REINFORCE algorithm, and \textit{r.} indicates a reproduced method. Our method (SubSearch) achieves state-of-the-art results among SFT-free RL agents.}
\label{tab:main_results}
\end{table*}

\header{SubSearch variants}
We experiment with both base and instruct versions of Qwen3.2-3B for SubSearch, and we observe a significant performance drop when training the instruct method with GRPO. As shown in Table~\ref{tab:base_vs_instruct}, StepSearch-base consistently outperforms across both general and multi-hop QA benchmarks.

\begin{table*}[h]
\centering
\small
\setlength{\tabcolsep}{1mm}
\resizebox{\textwidth}{!}{
\begin{tabular}{lccccccc}
\toprule
\textbf{Model} & \textbf{NQ$^\dagger$} & \textbf{TriviaQA$^\ast$} & \textbf{PopQA$^\ast$} & \textbf{HotpotQA$^\dagger$} & \textbf{2wiki$^\ast$} & \textbf{MuSiQue$^\ast$} & \textbf{Bamboogle$^\ast$} \\
\midrule
SubSearch-\textbf{base} & \textbf{0.463} & \textbf{0.597} & \textbf{0.435} & \textbf{0.362} & \textbf{0.351} & 0.101 & \textbf{0.263} \\
SubSearch-\textbf{instruct} & 0.325 & 0.523 & 0.353 & 0.300 & 0.294 & \textbf{0.110} & 0.219 \\
\bottomrule
\end{tabular}%
}
\caption{Performance comparison between base and instruct backbones for SubSearch.}
\label{tab:base_vs_instruct}
\end{table*}

\subsection{Ablation}

\begin{figure}[h]
    \centering
    \includegraphics[width=\textwidth]{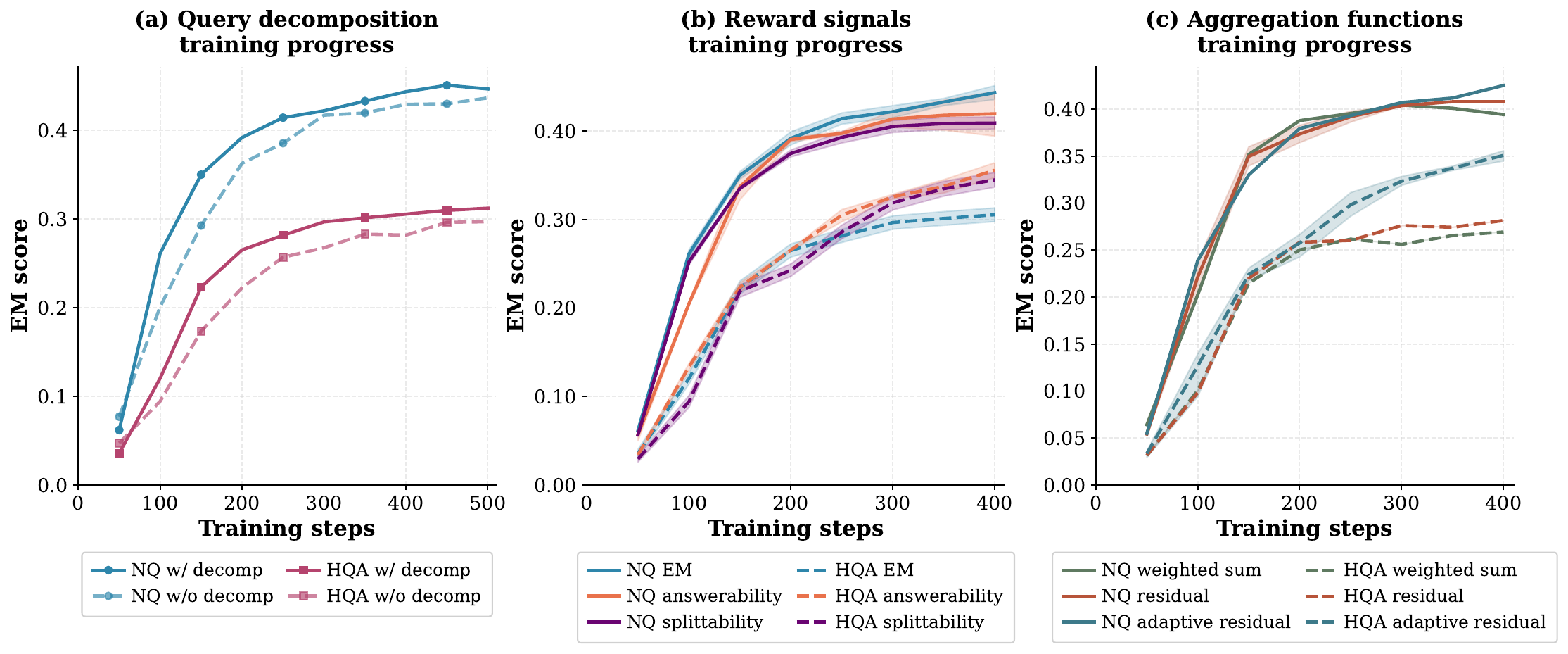}
    \caption{Training Progress of Qwen2.5-3B-base (a) with and without query decomposition via prompting on NQ and HotpotQA, (b) using  EM, answerability and decomposition as reward signals, and (c) using weighted sum, residual or adaptive residual reward aggregation.}
    \label{fig:training_lines}
\end{figure}

\begin{figure}[h]
    \centering
    \includegraphics[width=\textwidth]{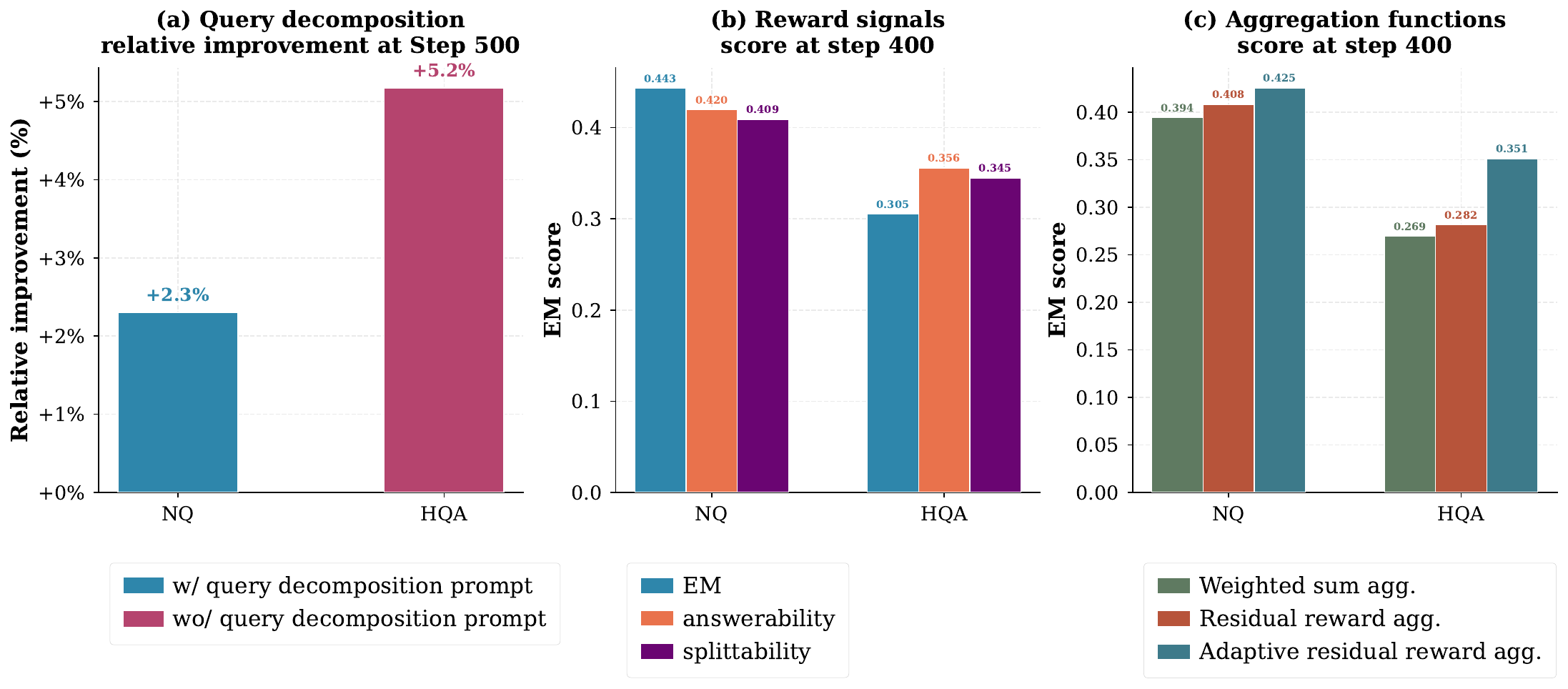}
    \caption{Relative improvements on Qwen2.5-3B-base (a) with and without query decomposition via prompting on NQ and HotpotQA, (b) using  EM, answerability and decomposition as reward signals, and (c) using weighted sum, residual or adaptive residual reward aggregation.}
    \label{fig:bars}
\end{figure}

\header{Effect of query decomposition}
%% make it clear it is prompt not reward
Figures \ref{fig:training_lines}(a) and \ref{fig:bars}(b) show the effect of query decomposition (via prompting) compared to query rewriting without decomposition. We can see a higher relative improvement on the HotpotQA dataset which inherently needs to reason and aggregate over multiple pieces of information, where we expect decomposition to be natural, while in the NQ dataset which contains more factoid queries, the relative improvement is smaller.

% \begin{figure}[h]
%     \centering
%     \includegraphics[width=\textwidth]{images/all_lines.pdf}
%     \caption{Training Progress of Qwen2.5-3B-base (a) with and without query decomposition via prompting on NQ and HotpotQA, (b) using  EM, answerability and splittability as reward signals, and (c) using weighted sum, residual or adaptive residual reward aggregation. \textcolor{orange}{[text in figures is way too small]}}
%     \label{fig:training_lines}
% \end{figure}

% \begin{figure}[h]
%     \centering
%     \includegraphics[width=\textwidth]{images/all_bars.pdf}
%     \caption{Relative improvements on Qwen2.5-3B-base (a) with and without query decomposition via prompting on NQ and HotpotQA, (b) using  EM, answerability and splittability as reward signals, and (c) using weighted sum, residual or adaptive residual reward aggregation. \textcolor{orange}{[text in figures is way too small]}}
%     \label{fig:bars}
% \end{figure}

% \begin{figure}[h]
%     \centering
%     \includegraphics[width=0.8\textwidth]{images/ablation_query_decomposition.pdf}
%     \caption{(a) Training Progress of Qwen2.5-3B-base with and without query decomposition via prompting on NQ and HotpotQA. (b) The relative improvement of query decomposition at step 500.}
%     \label{fig:query_decomposition}
% \end{figure}

\header{Reward variants} Figures \ref{fig:training_lines}(b) and \ref{fig:bars}(b) illustrate the effect of different types of reward on the NQ and HotpotQA dataset. We conducted this study to identify which intermediate training signal is better for generating a correct answer given a query. The EM reward refers to the original final-answer signal presented in Search-R1 \citep{jin2025search}, while answerability refers to the subquery-level reward in Eq.~\ref{eq:reward_answerability} and splitabilitty refers to the decomposition-level signal in Eq.~\ref{eq:reward_decomposition} We observe that in HotpotQA, both answerability and decomposition significantly improve the performance of the model, while answerability shows a small advantage. On the other hand, we see that intermediate reward training harms performance on NQ, a dataset where decomposition is not necessarily needed.

% \begin{figure}[h]
%     \centering
%     \includegraphics[width=0.8\textwidth]{images/rewards_ablation.pdf}
%     \caption{(a) Training Progress of Qwen2.5-3B-base with EM, answerability and splittability as reward signals. (b) The relative improvement of each reward signal at step 400.}
%     \label{fig:rewards}
% \end{figure}

\header{Aggregation functions} Figures \ref{fig:training_lines}(c) and \ref{fig:bars}(c) show different aggregation functions for integrating the intermediate reward signals with the final sparse EM reward. We tested with a simple weighted linear combination, however, it punishes reasoning traces for which the subquery generation is not optimal even if the model arrives at the correct answer. To avoid that, we moved to a residual (step-wise) reward function, where the model integrates the intermediate rewards only when it does not reach the correct answer. Finally, we add an adaptive weight that acts as a velocity, pushing the intermediate rewards more when the model has not reached the correct answer in the last steps, and vice versa. We describe them in Eq.~\ref{eq:aggregation_functions}, where we assume \(r_{\text{intermediate}}\) to represent any aggregation of intermediate rewards, as described in our setup in Eq. \ref{eq:adaptive_aggregation_simple}. We show the evolution of the adaptive beta \((\beta)\) during training in Figure \ref{fig:beta_evolution} in Appendix \ref{appendix:adaptive_beta}.
%
% \begin{figure}[h]
%     \centering
%     \includegraphics[width=0.8\textwidth]{images/aggregation_ablation.pdf}
%     \caption{(a) Training Progress of Qwen2.5-3B-base with EM, answerability and splittability as reward signals. (b) The relative improvement of each reward signal at step 400.}
%     \label{fig:agregation_functions}
% \end{figure}
%
\begin{align}
\begin{aligned}
\text{Weighted Sum:} \quad 
r &= \alpha \, r_{\text{answer}} + \beta \, r_{\text{intermediate}} \\
\text{Residual:} \quad 
r &= r_{\text{answer}} + \beta_t \, r_{\text{intermediate}} (1 - r_{\text{answer}}) \\
\text{Adaptive Residual:} \quad 
r &= r_{\text{answer}} + \beta_t \, r_{\text{intermediate}} (1 - r_{\text{answer}}), \quad \beta_t = f(t).
\end{aligned}
\label{eq:aggregation_functions}
\end{align}

\header{Format} Figure \ref{fig:format} in Appendix \ref{appendix:format} shows that adding formatting to the reasoning trace generation is vital to avoid early model collapse. While for non-decomposed generations the model seems more stable during training, adding parallel decomposition increases the complexity of the generation and therefore makes it more volatile. Hence, we use a combination of outcome-only, intermediate, and format rewards for stable and efficient training of SubSearch.

\section{Conclusion}

We have proposed SubSearch, a deep search agent that incentivizes robust step-by-step reasoning through intermediate reward signals at the subquery and decomposition level. Through query decomposition and by rewarding each query split, alongside how answerable each subquery is, SubSearch effectively learns to aggregate over a well-curated query reasoning tree. Experimental results show that SubSearch outperforms other SFT-Free methods on benchmarks such as NQ, TriviaQA, HotpotQA and 2wiki. While SubSearch integrates intrinsic process signals in the form of subquery-dependent rewards, it also adds computational complexity. Future work should explore how to make intrinsic intermediate reward calculation more efficient. Moreover, the added answerability reward is conditioned on the quality of the retriever, while only optimizing the reasoning agent and not the search engine. We aim to study the possibility of optimizing both the generator and retriever, with specialized signals for each module of the pipeline.

\bibliographystyle{colm2026_conference}
\bibliography{main}

\appendix
\section{Experimental Setup}\label{appendix:experimental_setup}

We follow an experimental setup similar to the one used in previous work \citep{jin2025search, sun2025zerosearchincentivizesearchcapability, zheng2025deepresearcher}, where we combine the Natural Questions (NQ) and HotpotQA datasets for training. We use Qwen2.5-3B and Qwen2.5-3B Instruct as backbone to our model and train using group relative policy optimization (GRPO) using the \textit{verl} framework. We apply GRPO with a group size of 5, a rollout temperature of 1.0, a training batch size of 512, and a validation batch size of 256. We set a maximum prompt length to 4096 tokens, a response length to 500 tokens, and an observation length of 1200 tokens. We employ a learning rate of $1 \times 10^{-6}$ and a KL divergence coefficient ($\text{kl\_loss\_coef}$) of $0.001$. The model is trained using 4 NVIDIA H100 GPUs.

\section{Format Analysis}
\label{appendix:format}

\begin{figure}[htbp]
    \centering
    \includegraphics[width=0.5\textwidth]{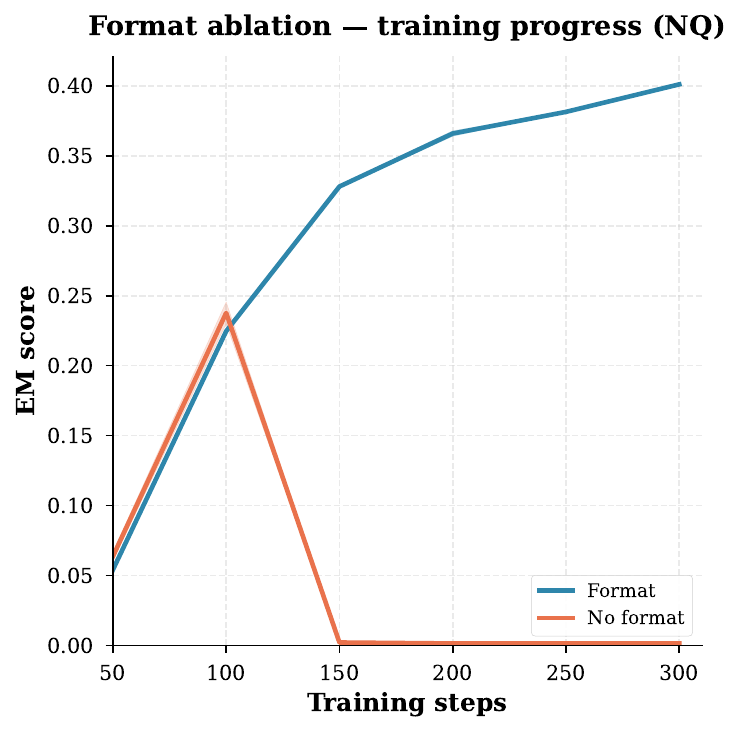}
    \caption{Training progress of Qwen2.5-3B-base on NQ with GRPO and EM as reward, with and without using a format signal.}
    \label{fig:format}
\end{figure}

\section{Query Decomposition and Rewards}
\label{appendix:decomposition_and_rewards}

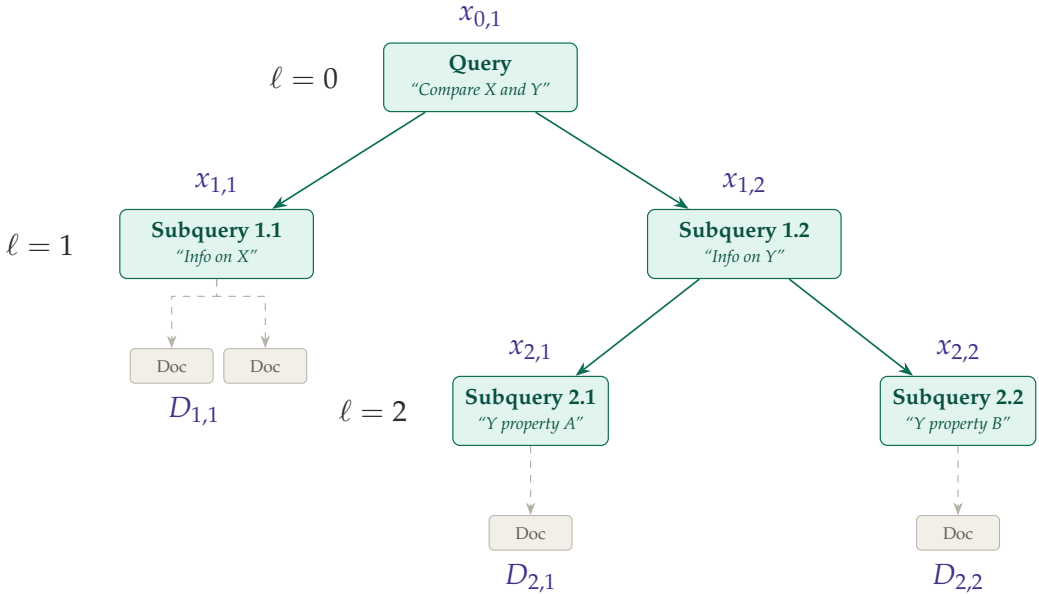
\begin{figure}[H]
  \centering
  \resizebox{\columnwidth}{!}{%
    %% ── TikZ styles ─────────────────────────────────────────────────
\tikzset{
  %% query box - WIDER
  qbox/.style={
    draw, rounded corners=3pt, inner sep=5pt,
    minimum width=2.8cm, minimum height=0.8cm,
    align=center, font=\small,
    fill=tealFill, draw=tealStroke, line width=0.5pt,
    text=tealText
  },
  %% document box - SLIGHTLY BIGGER
  dbox/.style={
    draw, rounded corners=2pt, inner sep=3pt,
    minimum width=1.2cm, minimum height=0.5cm,
    align=center, font=\scriptsize,
    fill=docFill, draw=docStroke, line width=0.4pt,
    text=docBody
  },
  %% layer label
  layerlabel/.style={
    font=\large\sffamily, text=grayText
  },
  %% notation label
  notationlabel/.style={
    font=\large\ttfamily, text=purpleText
  },
  %% arrow
  arr/.style={
    -Stealth, line width=0.7pt, tealStroke
  },
  %% dashed arrow to docs
  docarr/.style={
    -Stealth, line width=0.5pt, docStroke, dashed
  },
}

\begin{tikzpicture}[
  node distance=6pt and 10pt,
  level distance=1.4cm,
  sibling distance=3.0cm
]

%% ════════════════════════════════════════════════════════════════
%% LAYER 0 - Root query
%% ════════════════════════════════════════════════════════════════
\node[qbox] (q00) {%
  \textbf{Query}\\
  {\scriptsize\itshape``Compare X and Y''}
};

% Layer label
\node[layerlabel, left=15pt of q00] (L0) {\(\ell=0\)};

% Notation label
\node[notationlabel, above=2pt of q00] {\(x_{0,1}\)};

%% ════════════════════════════════════════════════════════════════
%% LAYER 1 - First decomposition
%% ════════════════════════════════════════════════════════════════
\node[qbox, below left=1.4cm and 1.0cm of q00] (q11) {%
  \textbf{Subquery 1.1}\\
  {\scriptsize\itshape``Info on X''}
};

\node[qbox, below right=1.4cm and 1.0cm of q00] (q12) {%
  \textbf{Subquery 1.2}\\
  {\scriptsize\itshape``Info on Y''}
};

% Layer label
\node[layerlabel, left=15pt of q11] (L1) {\(\ell=1\)};

% Notation labels
\node[notationlabel, above=2pt of q11] {\(x_{1,1}\)};
\node[notationlabel, above=2pt of q12] {\(x_{1,2}\)};

% Arrows from root
\draw[arr] (q00) -- (q11);
\draw[arr] (q00) -- (q12);

%% ════════════════════════════════════════════════════════════════
%% LAYER 2 - Second decomposition (only from q12)
%% ════════════════════════════════════════════════════════════════
\node[qbox, below left=1.4cm and 0.55cm of q12, minimum width=2.2cm] (q21) {%
  \textbf{Subquery 2.1}\\
  {\scriptsize\itshape``Y property A''}
};

\node[qbox, below right=1.4cm and 0.55cm of q12, minimum width=2.2cm] (q22) {%
  \textbf{Subquery 2.2}\\
  {\scriptsize\itshape``Y property B''}
};

% Layer label
\node[layerlabel, left=15pt of q21] (L2) {\(\ell=2\)};

% Notation labels
\node[notationlabel, above=2pt of q21] {\(x_{2,1}\)};
\node[notationlabel, above=2pt of q22] {\(x_{2,2}\)};

% Arrows from q12
\draw[arr] (q12) -- (q21);
\draw[arr] (q12) -- (q22);

%% ════════════════════════════════════════════════════════════════
%% OPTIONAL: Documents (for q11, q21, q22)
%% ════════════════════════════════════════════════════════════════
% Documents for q11
\node[dbox, below=1.0cm of q11, xshift=-0.65cm] (d111) {Doc};
\node[dbox, right=4pt of d111] (d112) {Doc};

\node[notationlabel, below=2pt of d111, xshift=0.32cm] {\(D_{1,1}\)};

\draw[docarr] (q11.south) -- ++(0,-0.25cm) -| (d111.north);
\draw[docarr] (q11.south) -- ++(0,-0.25cm) -| (d112.north);

% Documents for q21
\node[dbox, below=1.0cm of q21] (d211) {Doc};
\node[notationlabel, below=2pt of d211] {\(D_{2,1}\)};
\draw[docarr] (q21) -- (d211);

% Documents for q22
\node[dbox, below=1.0cm of q22] (d221) {Doc};
\node[notationlabel, below=2pt of d221] {\(D_{2,2}\)};
\draw[docarr] (q22) -- (d221);

%% ════════════════════════════════════════════════════════════════
%% LEGEND / EXPLANATION
%% ════════════════════════════════════════════════════════════════
% \node[below=2.4cm of q00, font=\scriptsize, text=grayText, align=center] (legend) {%
%   \textbf{Notation:} \(\ell\): layer (root=0), \(i\): index within layer, 
%   \(x_{\ell,i}\): subquery, \(D_{\ell,i}\): documents
% };

%% ════════════════════════════════════════════════════════════════
%% ANNOTATIONS - Show the structure
%% ════════════════════════════════════════════════════════════════
% % Annotation for layer 1 decomposition
% \node[below=0.12cm of q11, xshift=1.4cm, font=\scriptsize\itshape, text=grayText] 
%     {\(n=2\) subqueries};

% % Annotation for layer 2 decomposition  
% \node[below=0.12cm of q21, xshift=1.1cm, font=\scriptsize\itshape, text=grayText] 
%     {\(n=2\) subqueries};

\end{tikzpicture}%
  }
  \caption{SubSearch notation for query decomposition and intermediate rewards computation. \(l\) denotes the decomposition layer (root = 0), \(i\) denotes the index within each layer, \(x_{l,i}\) represents a subquery at layer \(l\) and index \(i\), and \(D_{l,i}\) represents the retrieved documents for subquery \(x_{l,i}\).}
  \label{fig:query_decomp_and_rewards}
\end{figure}

% \node[below=4cm of q00, font=\tiny, text=grayText, align=left] (legend) {%
%   \textbf{Notation:} \(\ell\): layer (root=0), \(i\): index, 
%   \(x_{\ell,i}\): subquery, \(D_{\ell,i}\): documents
% };

\section{Adaptive beta}
\label{appendix:adaptive_beta}

\begin{figure}[htbp]
    \centering
    \includegraphics[width=0.5\textwidth]{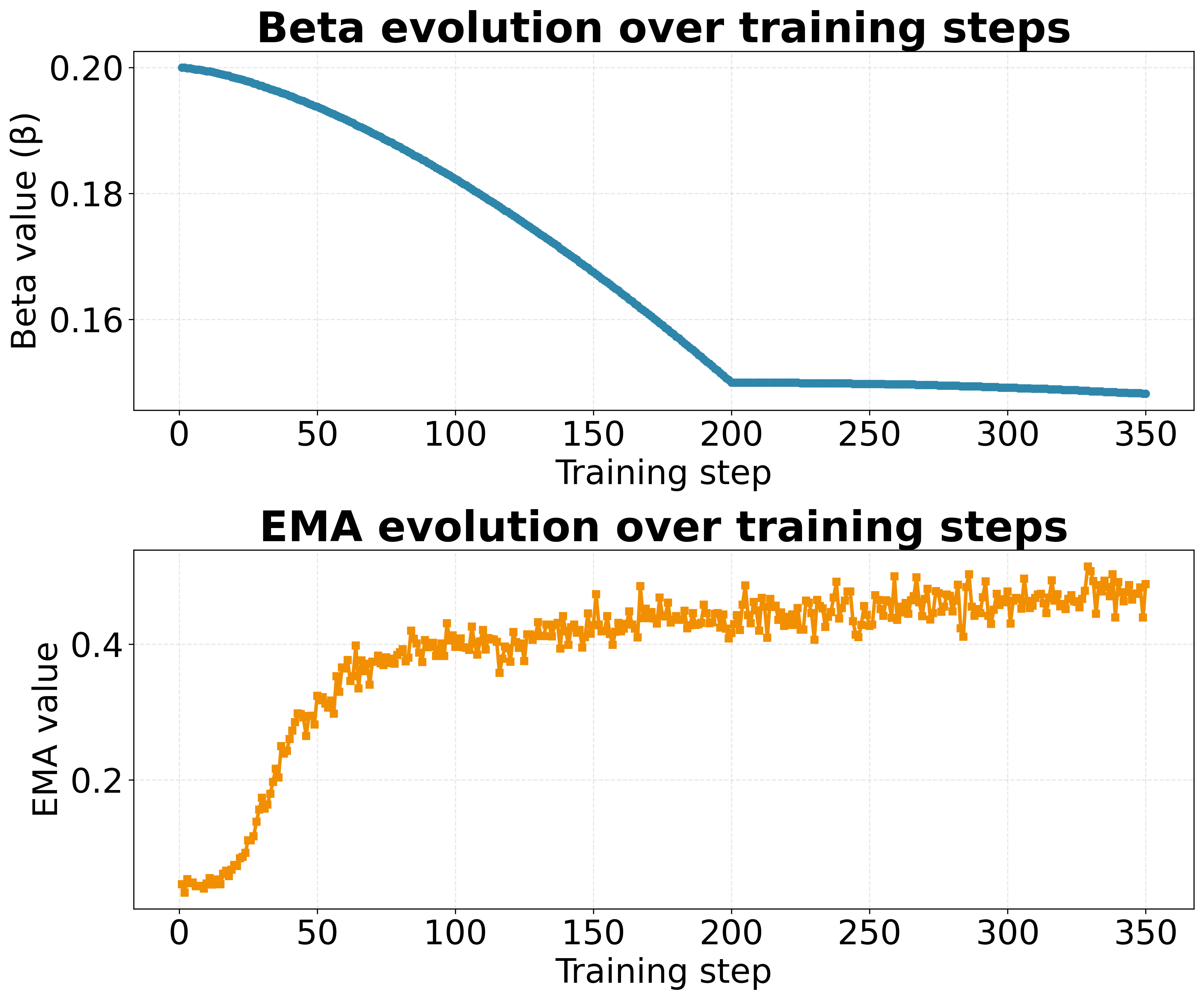}
    \caption{Adaptive beta \((\beta)\) evolution over training steps. A higher value highlights more weight to the intermediate rewards, while a lower value shows the model becomes better at giving a correct answer and therefore gives priority to the binary outcome reward. Higher EMA highlights giving more weight to recent reasoning traces compared to older ones.}
    \label{fig:beta_evolution}
\end{figure}

\section{Examples}
We  include three case studies to illustrate how SubSearch successfully answer questions (Table~\ref{table:use-case-1}, \ref{table:use-case-2}, \ref{table:use-case-3}) and one case where SubSearch is not successful due to a failure in decomposition (Table~\ref{table:use-case-4}),

\begin{table*}[t]
\small
\begin{tabular}{p{0.98\textwidth}}

\hline
\textbf{Question:} Which bank has more branches, China CITIC Bank or UniCredit? \\
\hline
\textbf{Ground Truth:} UniCredit \\
\hline

\textcolor{thinkcolor}{\textbf{\textless think\textgreater}} First, I need to find information about China CITIC Bank. \textcolor{thinkcolor}{\textbf{\textless/think\textgreater}} \\

\textcolor{searchcolor}{\textbf{\textless search\textgreater}} how many branches does UniCredit have bank \textcolor{searchcolor}{\textbf{\textless/search\textgreater}} \\

\textcolor{infocolor}{\textbf{\textless information\textgreater}} Doc 1(Title: "UniCredit Bank Romania") UniCredit Bank Romania UniCredit Bank is a leading European Bank, member of UniCredit Group, with a network of 8.500 branches in 17 European countries and it is present in another 50 international markets Being one of the top 5 banks on the Romanian market, UniCredit Bank has 208 branches and almost 3000 employees. At the end of 2015, the bank had total assets in amount of 34,6 billion RON and included approximately 600.000 active customers. Among the offered services are: financial solutions for individuals, SMEs, companies and freelancers. UniCredit Bank is recognized for specializing in the real estate market and.. Doc 2(Title: UniCredit) UniCredit UniCredit S.p.A. is an Italian global banking and financial services company. Its network spans 50 markets in 17 countries, with more than 8,500 branches and over 147,000 employees. Its strategic position in Western and Eastern Europe gives the group one of the region's highest market shares. The company is a component of the Euro Stoxx 50 stock market index. UniCredit Group was the outcome of the 1998 merger of several Italian banking groups, which the majority one were Unicredito (banks from Turin, Verona and Treviso) and Credito Italiano (consists of Rolo Banca, Banca Popolare di Rieti), hence the name.. Doc 3(Title: UniCredit) subsidiary Bank Austria). Until 2018 UniCredit was one of the UEFA Champions League's main sponsors. UniCredit UniCredit S.p.A. is an Italian global banking and financial services company. Its network spans 50 markets in 17 countries, with more than 8,500 branches and over 147,000 employees. Its strategic position in Western and Eastern Europe gives the group one of the region's highest market shares. The company is a component of the Euro Stoxx 50 stock market index. UniCredit Group was the outcome of the 1998 merger of several Italian banking groups, which the majority one were.. \textcolor{infocolor}{\textbf{\textless/information\textgreater}} \\

\textcolor{thinkcolor}{\textbf{\textless think\textgreater}} Based on the information found, I need to compare the number of branches of China CITIC Bank and UniCredit. \textcolor{thinkcolor}{\textbf{\textless/think\textgreater}} \\

\textcolor{searchcolor}{\textbf{\textless search\textgreater}} how many branches does China CITIC Bank have \textcolor{searchcolor}{\textbf{\textless/search\textgreater}} \\

\textcolor{infocolor}{\textbf{\textless information\textgreater}} Doc 1(Title: "China CITIC Bank") China CITIC Bank China CITIC Bank () is China’s seventh-largest lender in terms of total assets. It was known as CITIC Industrial Bank until it changed its name in August 2005. China CITIC Bank, established in 1987, is a nationally comprehensive and internationally oriented commercial bank. The bank operates in almost 130 countries, and maintains a strong foothold on the mainland banking industry. The bank operates 78 branches in the mainland, and 622 sub-branches, located in economically developed regions of China. In total, there are 773 branch offices in China, as of Q4 2011. In 1984, the chairman of CITIC.. Doc 2(Title: "China CITIC Bank") financing services for small enterprises, trading services, cash management, asset management, international settlement and trade financing, capital market business, including foreign exchange businesses, bond businesses, trading of financing and derivative products, financial service platforms and financial solutions. China CITIC Bank China CITIC Bank () is China’s seventh-largest lender in terms of total assets. It was known as CITIC Industrial Bank until it changed its name in August 2005. China CITIC Bank, established in 1987, is a nationally comprehensive and internationally oriented commercial bank. The bank operates in almost 130 countries, and maintains a strong foothold on the mainland banking industry.. Doc 3(Title: UniCredit) the bank was also relocated from Genoa (inherited from Credito Italiano) to 17 via Minghetti, Rome. In the same year, two more acquisitions were carried out: ATF Bank, which ranks fifth out of domestic banks in Kazakhstan with 154 branches, and Ukrsotsbank, a universal bank in Ukraine. With these two banks the Group extended its operations in this area to 19 countries (including Central Asia). However, in November 2012, Kazakh government sources declared UniCredit is in talks with Kazakh investors over the sale of a controlling stake in ATF Bank. In 2010, UniCredit S.p.A. absorbed its Italian banking subsidiaries: UniCredit.. \textcolor{infocolor}{\textbf{\textless/information\textgreater}} \\

\textcolor{thinkcolor}{\textbf{\textless think\textgreater}} Based on the information, China CITIC Bank has 773 branches, while UniCredit has 8,500 branches. \textcolor{thinkcolor}{\textbf{\textless/think\textgreater}} \\

\textcolor{answercolor}{\textbf{\textless answer\textgreater}} UniCredit \textcolor{answercolor}{\textbf{\textless/answer\textgreater}} \\

\hline
\end{tabular}\caption{SubSearch case study 1 (successful): SubSearch can decompose the initial information need into parallel subqueries and aggregate the retrieved documents to answer correctly.}
\label{table:use-case-1}
\end{table*}

\begin{table*}[t]
\small
\begin{tabular}{p{0.98\textwidth}}

\hline
\textbf{Question:} Which team did Joe Buck's father broadcast for? \\
\hline
\textbf{Ground Truth:} St. Louis Cardinals \\
\hline

\textcolor{thinkcolor}{\textbf{\textless think\textgreater}} First, I need to find information about Joe Buck. \textcolor{thinkcolor}{\textbf{\textless/think\textgreater}} \\

\textcolor{searchcolor}{\textbf{\textless search\textgreater}} who is joe buck father broadcast \textcolor{searchcolor}{\textbf{\textless/search\textgreater}} \\

\textcolor{infocolor}{\textbf{\textless information\textgreater}} Doc 1(Title: "Joe Buck") but rather to vocal cord paralysis likely caused by anesthesia used during multiple hair transplantation procedures. From 1993 to 2011, Buck was married to Ann Archambault, with whom he had two children. He married NFL Network and now ESPN reporter and former Bronco cheerleader Michelle Beisner on April 12, 2014. On April 26, 2018 they welcomed twin sons Wyatt Joseph and Blake Andrew. Joe Buck Joseph Francis Buck (born April 25, 1969) is an American sportscaster and the son of sportscaster Jack Buck. He has won numerous Sports Emmy Awards for his work with Fox Sports, including his roles as.. Doc 2(Title: "Joe Buck") the team's broadcasting crew. In 1994, Buck was hired by Fox, and at the age of 25 became the youngest man ever to announce a regular slate of National Football League games on network television. In , he was named Fox's lead play-by-play voice for Major League Baseball, teaming with Tim McCarver, who had previously worked with his father on CBS. That year, he became the youngest man to do a national broadcast (for all nine innings and games, as a network employee as opposed to simply being a representative of one of the participating teams) for a World Series.. Doc 3(Title: "Joe Buck") Joe Buck Joseph Francis Buck (born April 25, 1969) is an American sportscaster and the son of sportscaster Jack Buck. He has won numerous Sports Emmy Awards for his work with Fox Sports, including his roles as lead play-by-play announcer for the network's National Football League and Major League Baseball coverage, and is a three-time recipient of the National Sportscaster of the Year award. Since 1996, he has served as the play-by-play announcer for the World Series, each year, with the exceptions of 1997 and 1999. Since 2015, he's hosted ""Undeniable with Joe Buck"" on Audience Network. Buck was born\textcolor{infocolor}{\textbf{\textless/information\textgreater}} \\

\textcolor{thinkcolor}{\textbf{\textless think\textgreater}} I now need to find which team his father broadcast for. \textcolor{thinkcolor}{\textbf{\textless/think\textgreater}} \\

\textcolor{searchcolor}{\textbf{\textless search\textgreater}} who did jack buck broadcast for \textcolor{searchcolor}{\textbf{\textless/search\textgreater}} \\

\textcolor{infocolor}{\textbf{\textless information\textgreater}} Doc 1(Title: "Jack Buck") a railroad accountant who commuted weekly to New Jersey. From an early age, Buck dreamed of becoming a sports announcer with his early exposure to sports broadcasting coming from listening to Boston Red Sox baseball games announced by Fred Hoey. Part of his childhood coincided with the Great Depression, and Buck remembered his family sometimes using a metal slug to keep a coin-operated gas meter going during the winter to provide heat for their home. In 1939, his family moved to Cleveland, Ohio to join their father, who had a job with the Erie Railroad. Soon after though, Buck's father.. Doc 2(Title: "Jack Buck") after vomiting on a plane during a flight after a game, and was out for a considerable amount of time. While Verne Lundquist replaced Summerall on games with lead analyst John Madden, Buck (who was at the time the network's lead Major League Baseball announcer) filled in for Lundquist, teaming with Dan Fouts to call two games (both of which coincidentally featured the Cardinals, who had moved from St. Louis to Arizona by that time). While much better known for his baseball and football commentary, Jack Buck was also the original voice of the St. Louis Blues of the National.. Doc 3(Title: "Jack Buck") Jack Buck John Francis ""Jack"" Buck (August 21, 1924 – June 18, 2002) was an American sportscaster, best known for his work announcing Major League Baseball games of the St. Louis Cardinals. His play-by-play work earned him recognition from numerous Halls of Fame, such as the National Baseball Hall of Fame, the Pro Football Hall of Fame, and the National Radio Hall of Fame. He has also been inducted as a member of the St. Louis Cardinals Hall of Fame Museum. Buck was born in Holyoke, Massachusetts, the third of seven children of Earle and Kathleen Buck.. \textcolor{infocolor}{\textbf{\textless/information\textgreater}} \\

\textcolor{thinkcolor}{\textbf{\textless think\textgreater}} Therefore, Joe Buck's father broadcast for the St. Louis Cardinals. \textcolor{thinkcolor}{\textbf{\textless/think\textgreater}} \\

\textcolor{answercolor}{\textbf{\textless answer\textgreater}} St. Louis Cardinals \textcolor{answercolor}{\textbf{\textless/answer\textgreater}} \\

\hline
\end{tabular}\caption{SubSearch case study 2 (successful): SubSearch can decompose the initial information need into sequential subqueries (we need the answer of one to formulate the other) and aggregate the retrieved documents to answer correctly.}
\label{table:use-case-2}
\end{table*}

\begin{table*}[t]
\small
\begin{tabular}{p{0.98\textwidth}}

\hline
\textbf{Question:} In what year did the man who shot Chris Stockley, of The Dingoes, die? \\
\hline
\textbf{Ground Truth:} 1987 \\
\hline

\textcolor{thinkcolor}{\textbf{\textless think\textgreater}} First, I need to find information about Chris Stockley. \textcolor{thinkcolor}{\textbf{\textless/think\textgreater}} \\

\textcolor{searchcolor}{\textbf{\textless search\textgreater}} when did Chris Stockley of The Dingoes die \textcolor{searchcolor}{\textbf{\textless/search\textgreater}} \\

\textcolor{infocolor}{\textbf{\textless information\textgreater}} Doc 1(Title: "The Dingoes") The Dingoes The Dingoes is an Australian country rock band initially active from 1973 to 1979, formed in Melbourne which relocated to the United States from 1976. Most stable line-up was John Bois on bass guitar, John Lee on drums, Broderick Smith on vocals and harmonica, Chris Stockley on guitar, and Kerryn Tolhurst on guitar. Mal Logan (who provided keyboards on the first LP) on keyboards joined after Stockley was hospitalised when shot in the stomach by Melbourne drug-dealer, Dennis Allen, who was attempting to gate crash a party. The Dingoes' debut single, ""Way Out West"", was released in November.. Doc 2(Title: "The Dingoes") and Tolhurst although Bois' book confirms that the song was written by Tolhurst alone. A week before the single was released Stockley received a serious gunshot wound during an incident at a party in Melbourne that resulted in a two-month stay in hospital, initially described as an 'accidental shooting', according to music historian, Ian McFarlane's ""Encyclopedia of Australian Rock and Pop"",...\textcolor{infocolor}{\textbf{\textless/information\textgreater}} \\

\textcolor{thinkcolor}{\textbf{\textless think\textgreater}} Now, I need to find information about the man who shot him. \textcolor{thinkcolor}{\textbf{\textless/think\textgreater}} \\

\textcolor{searchcolor}{\textbf{\textless search\textgreater}} who shot Chris Stockley of The Dingoes \textcolor{searchcolor}{\textbf{\textless/search\textgreater}} \\

\textcolor{infocolor}{\textbf{\textless information\textgreater}} Doc 1(Title: "The Dingoes") and Tolhurst although Bois' book confirms that the song was written by Tolhurst alone. A week before the single was released Stockley received a serious gunshot wound during an incident at a party in Melbourne that resulted in a two-month stay in hospital, initially described as an 'accidental shooting', according to music historian, Ian McFarlane's ""Encyclopedia of Australian Rock and Pop"", Stockley was shot by notorious drug dealer Dennis Allen, who was trying to gate crash the party. An eight-hour benefit concert was held for Stockley on 4 November 1973 at Leggett's Ballroom, Greville Street, Prahran. While recuperating, Stockley was.. Doc 2(Title: "The Dingoes") The Dingoes The Dingoes is an Australian country rock band initially active from 1973 to 1979, formed in Melbourne which relocated to the United States from 1976. Most stable line-up was John Bois on bass guitar, John Lee on drums, Broderick Smith on vocals and harmonica, Chris Stockley on guitar, and Kerryn Tolhurst on guitar... \textcolor{infocolor}{\textbf{\textless/information\textgreater}} \\

\textcolor{thinkcolor}{\textbf{\textless think\textgreater}} I now need to find when Dennis Allen died. \textcolor{thinkcolor}{\textbf{\textless/think\textgreater}} \\

\textcolor{searchcolor}{\textbf{\textless search\textgreater}} when did Dennis Allen die \textcolor{searchcolor}{\textbf{\textless/search\textgreater}} \\

\textcolor{infocolor}{\textbf{\textless information\textgreater}} Doc 1(Title: "The Dingoes") and Tolhurst although Bois' book confirms that the song was written by Tolhurst alone. A week before the single was released Stockley received a serious gunshot wound during an incident at a party in Melbourne that resulted in a two-month stay in hospital, initially described as an 'accidental shooting', according to music historian, Ian McFarlane's ""Encyclopedia of Australian Rock and Pop"", Stockley was shot by notorious drug dealer Dennis Allen, who was trying to gate crash the party. An eight-hour benefit concert was held for Stockley on 4 November 1973 at Leggett's Ballroom, Greville Street, Prahran. While recuperating, Stockley was.. Doc 2(Title: "Dennis Allen (criminal)") in 1985. One victim who survived was guitarist Chris Stockley of The Dingoes, whom Allen shot in the stomach while attempting to gatecrash a party. Allen received a ten-year prison sentence for rape during the 1970s. It is also reported that he was a major drug dealer in the Richmond and South Yarra areas of Melbourne during the 1980s. New South Wales Police Detective-Sergeant Roger Rogerson was convicted of supplying heroin in a deal with Allen,...\textcolor{infocolor}{\textbf{\textless/information\textgreater}} \\

\textcolor{thinkcolor}{\textbf{\textless think\textgreater}} Therefore, the man who shot Chris Stockley died in 1987. \textcolor{thinkcolor}{\textbf{\textless/think\textgreater}} \\

\textcolor{answercolor}{\textbf{\textless answer\textgreater}} 1987 \textcolor{answercolor}{\textbf{\textless/answer\textgreater}} \\

\hline
\end{tabular}\caption{SubSearch case study 3 (successful): SubSearch can decompose the initial information need into sequential subqueries (we need the answer of one to formulate the other) and aggregate the retrieved documents to answer correctly. However, compared to case study 2, we also see the model generated a search that is useless (see subquery 1).}
\label{table:use-case-3}
\end{table*}

\begin{table*}[t]
\small
\begin{tabular}{p{0.98\textwidth}}
\textbf{Case study: failed reasoning due to missing decomposition and spurious grounding} \\

\hline
\textbf{Question:} At what theater is the composer and lyricist for the musical Big Fish a residential artist? \\
\hline
\textbf{Ground Truth:} Ars Nova Theater \\
\hline

\textcolor{thinkcolor}{\textbf{\textless think\textgreater}} First, I need to find the composer and lyricist of Big Fish. \textcolor{thinkcolor}{\textbf{\textless/think\textgreater}} \\

\textcolor{searchcolor}{\textbf{\textless search\textgreater}} what is the theater of Big Fish musical composer lyricist residential artist \textcolor{searchcolor}{\textbf{\textless/search\textgreater}} \\

\textcolor{infocolor}{\textbf{\textless information\textgreater}} Doc 1(Title: "Big Fish (musical)") Big Fish (musical) Big Fish is a musical with music and lyrics by Andrew Lippa and book by John August. It is based on Daniel Wallace’s 1998 novel, """", and the 2003 film ""Big Fish"" written by John August and directed by Tim Burton. ""Big Fish"" revolves around the relationship between Edward Bloom, a travelling salesman, and his adult son Will, who looks for what is behind his father’s tall stories. The story shifts between two timelines. In the present-day real world, sixty-year-old Edward Bloom faces his mortality while Will prepares to become a father himself. In the storybook past.. Doc 2(Title: "Big Fish (musical)") 'Dirty Rotten Scoundrels.' His ingratiating singing and dancing bespeak the rare man comfortable in his own skin, and he has that indefinable charismatic spark that defines a star. ""Variety"" reviewer Marilyn Stasio wrote: ""Norbert Leo Butz is cutting loose in another one of his don’t-dare-miss-this perfs in 'Big Fish,' a show that speaks to anyone pining for a studiously heart-warming musical about the efforts of a dying man to justify a lifetime of lousy parenting to his alienated son."" Big Fish (musical) Big Fish is a musical with music and lyrics by Andrew Lippa and book by John August. It.. Doc 3(Title: "Big Fish: A Novel of Mythic Proportions") Big Fish: A Novel of Mythic Proportions Big Fish: A Novel of Mythic Proportions is a 1998 novel by Daniel Wallace. It was adapted into a film, ""Big Fish"", in 2003 by Tim Burton. A musical adaptation starring Norbert Leo Butz premiered in Chicago in April 2013. A young man (William Bloom), at the deathbed of his father (Edward Bloom), tries to reconcile his memories of his dad with the person he really is. Whereas he always saw his father as an irresponsible liar, he comes to understand his dad's exaggerations and their roots in reality. The book is written.. \textcolor{infocolor}{\textbf{\textless/information\textgreater}} \\

\textcolor{thinkcolor}{\textbf{\textless think\textgreater}} Now I need to find where Andrew Lippa is a resident artist. \textcolor{thinkcolor}{\textbf{\textless/think\textgreater}} \\

\textcolor{searchcolor}{\textbf{\textless search\textgreater}} where is the theater of composer lyricist Big Fish residential artist \textcolor{searchcolor}{\textbf{\textless/search\textgreater}} \\

\textcolor{infocolor}{\textbf{\textless information\textgreater}} Doc 1(Title: "Big Fish (musical)") Big Fish (musical) Big Fish is a musical with music and lyrics by Andrew Lippa and book by John August. It is based on Daniel Wallace’s 1998 novel, """", and the 2003 film ""Big Fish"" written by John August and directed by Tim Burton. ""Big Fish"" revolves around the relationship between Edward Bloom, a travelling salesman, and his adult son Will, who looks for what is behind his father’s tall stories. The story shifts between two timelines. In the present-day real world, sixty-year-old Edward Bloom faces his mortality while Will prepares to become a father himself. In the storybook past.. Doc 2(Title: "Big Fish (musical)") as Karl the Giant, Brenden Lovett as Amos Calloway/Doctor Bennet, Zoe Ioannou as the Girl in the Water, Zachary Webster as the Frustrated Fisherman, and Sam Wood and Brendan Godwin sharing the role of Young Will. The production was the 12 Chairs version, produced by RPG Productions and directed by Tyran Parke. The show received its London premiere in November 2017 at The Other Palace, starring Kelsey Grammer, Jamie Muscato and Matthew Seadon-Young. Sources: PlaybillVault; Internet Broadway database ""Big Fish"" was orchestrated by Larry Hochman, with additional orchestrations by Bruce Coughlin, for a fourteen-piece orchestra. The instrumentation calls for two.. Doc 3(Title: "Andrew Lippa") shortly. The recording was released on Decca Broadway and the vocal selections are available from Hal Leonard. ""Big Fish"", Lippa’s newest musical, with a book by John August, based on the 2003 film and the , premiered in Chicago, Illinois at the Oriental Theatre from April 2013 through May 5, 2013. It opened on Broadway at the Neil Simon Theatre October 6, 2013, following previews that began on September 5, 2013. Susan Stroman directed and choreographed, with scenic design by Julian Crouch, costume design by William Ivey Long and lighting design by Donald Holder. Norbert Leo Butz starred as Edward \textcolor{infocolor}{\textbf{\textless/information\textgreater}} \\

\textcolor{thinkcolor}{\textbf{\textless think\textgreater}} The musical is associated with the Neil Simon Theatre, so the composer is likely a resident artist there. \textcolor{thinkcolor}{\textbf{\textless/think\textgreater}} \\

\textcolor{answercolor}{\textbf{\textless answer\textgreater}} Neil Simon Theatre \textcolor{answercolor}{\textbf{\textless/answer\textgreater}} \\

\hline
\end{tabular}\caption{SubSearch case study 4 (unsuccessful): SubSearch gets stuck into generating the similar queries every time, finally reaching an answer that is not correct. While the reasoning goes well, the model fails in asking the correct 2nd question.}
\label{table:use-case-4}
\end{table*}

\end{document}